\documentclass{emulateapj}        

\def\etal{{\it et~al. }}

\newcommand{\am}[2]{$#1'\,\hspace{-1.7mm}.\hspace{.0mm}#2$}
\def\approxlt{\lower.2em\hbox{$\buildrel < \over \sim$}}
\def\approxgt{\lower.2em\hbox{$\buildrel > \over \sim$}}

\newcommand{\HI}{\mbox{H\,{\sc i}}}

\newcommand{\HII}{\mbox{H\,{\sc ii}}}
\newcommand{\IHI}{\mbox{${I}_{HI}$}}

\newcommand{\MHI}{\mbox{${M}_{HI}$}}

\def\la{\mathrel{\hbox{\rlap{\hbox{\lower4pt\hbox{$\sim$}}}\hbox{$<$}}}}
\def\ga{\mathrel{\hbox{\rlap{\hbox{\lower4pt\hbox{$\sim$}}}\hbox{$>$}}}}

\begin{document}

\title{The Photometric and Kinematic Structure of Face-On Disk
Galaxies. I. Sample Definition, H$\alpha$ Integral Field Spectroscopy,
and \HI~Line-Widths}
\shorttitle{H$\alpha$ IFS and \HI~Line-Widths of Face-On Disk Galaxies}

\author{David R. Andersen\altaffilmark{1,2}}

\affil{NRC Herzberg Institute of Astrophysics, 5071 W Saanich Road,
Victoria, BC V9E 2E7}
\email{david.andersen@cnrc-nrc.gc.ca}

\author{Matthew A. Bershady, Linda S. Sparke, John S. Gallagher, III,
Eric M. Wilcots} 

\affil{Department of Astronomy, University of Wisconsin, 475 N Charter
Street, Madison, WI 53706; mab@astro.wisc.edu, sparke@astro.wisc.edu,
jsg@astro.wisc.edu, ewilcots@astro.wisc.edu}

\author{Wim van Driel}

\affil{Observatoire de Paris, Section de Meudon, GEPI,
CNRS UMR 8111 and Universit\'e de Paris 7, 5 place Jules Janssen,
92195 Meudon 
Cedex, France; Wim.vanDriel@obspm.fr}

\author{Delphine Monnier-Ragaigne} 

\affil{Laboratoire de l'Acc\'el\'erateur Lin\'eaire, Universit\'e
Paris-Sud B\^atiment 200, BP 34, 91898 Orsay Cedex, France;
monnier@lal.in2p3.fr}

\altaffiltext{1}{Max Planck Institute for Astronomy, K\"onigstuhl 17, D-69117
Heidelberg}
\altaffiltext{2}{Visiting Astronomer, Kitt Peak National Observatory.
KPNO is operated by AURA, Inc.\ under contract to the National Science
Foundation.}

\begin{abstract}

We present a survey of the photometric and kinematic properties of 39
nearby, nearly face--on disk galaxies. Our approach exploits
echelle-resolution integral-field spectroscopy of the H$\alpha$
regions, obtained with DensePak on the WIYN 3.5m telescope Bench
Spectrograph. This data is complemented by \HI\ line-profiles observed
with the Nan\c{c}ay radio telescope for 25 of these sample galaxies.
Twelve additional line-widths are available for sample galaxies from
the literature. In this paper, we introduce the goals of this survey,
define the sample selection algorithm, and amass the integral field
spectroscopic data and \HI~line-widths. We establish
spatially-integrated H$\alpha$ line-widths for the sample. We test the
veracity of these spatially-integrated line profiles by convolving
narrow-band imaging data with velocity field information for one of
the sample galaxies, PGC 38268, and also by comparing to \HI\ line
profiles. We find \HI~ and H$\alpha$ line profiles to be similar in
width but different in shape, indicating we are observing different
spatial distributions of ionized and neutral gas in largely
axisymmetric systems with flat outer rotation-curves.  We also find
vertical velocity dispersions of the ionized disk gas within several
disk scale-lengths have a median value of 18 km s$^{-1}$ and an 80\%
range of 12-26 km s$^{-1}$. This is only a factor of $\sim$2 larger
than what is observed for neutral atomic and molecular gas. With
standard assumptions for intrinsic and thermal broadening for
H$\alpha$, this translates into a factor of three range in turbulent
velocities, between 8 and 25 km s$^{-1}$.

\end{abstract}

\keywords{galaxies: structure; galaxies: kinematics; galaxies: spiral}

\section{Introduction}

Two fundamental and intertwined cosmological questions are how
galaxies form and come to have their present structure. The current
approach to answering these questions involves comparing numerical and
semi-analytical structure-formation models to a suite of critical
observations of low and high redshift galaxies (e.g., Navarro, Frenk,
\& White 1996, 1997; van den Bosch 2000). For nearby disk galaxies one
of the salient observations which tests and constrains
structure-formation models is a measurement of halo shape.

An increasing number of cosmological simulations show that dark matter
halos are naturally triaxial (Davis \etal 1985; Frenk \etal 1988;
Dubinski \& Carlberg 1991; Jing \etal 1995; Jing \& Suto 2002), but
other physical processes, such as the dissipative infall of gas during
formation, may transform the halo into an oblate spheroid (Dubinski
1994). If dark matter halos are non-spherical, then this should have
an observational impact on the luminous matter in disks. Theory
predicts a number of observational consequences including effects on
(i) the orbits in lopsided galaxies (Levine \& Sparke 1998; Jog 1999;
Noordermeer, Sparke, Levine 2001; Jog 2002); (ii) the formation and
structure of bars and oval distortions (Athanassoula 2003); and (iii)
non-circular motions in dark matter dominated low surface brightness
(LSB) galaxies (Hayashi \etal 2004). Triaxial halos also should beget
elliptical disks (Franx \& de Zeeuw 1992, Jog 2000). Because disk
asymmetries may be linked to the shape of the halo, it is interesting
to measure deviations from axisymmetry. Indeed, many galaxies appear
to have lopsided light distributions (e.g.,
Zaritsky \& Rix 1997; Kornreich, Haynes, Lovelace 1998; Conselice,
Bershady, Jangren 2000; Bournaud et al. 2005), 
\HI~distributions (e.g. Baldwin, Lynden-Bell, Sancisi
1980), integrated \HI~line-profiles (Richter \&
Sancisi 1994), or kinematic features (e.g., Richter \&
Sancisi 1994; Swaters \etal 1999; Kornreich \etal 2000, Chemin \etal
2006), especially in
LSB galaxies (Swaters \etal 2003; Simon \etal 2003). However, a direct
measurement of disk ellipticity is challenging because it is difficult
to disentangle the intrinsic shape of galaxy disks from both
projection effects and spiral structure (e.g.  Rix \& Zaritsky 1995;
Schoenmakers \etal 1997, Schoenmakers 1999; Andersen \etal 2001;
Barnes \& Sellwood 2003). If better estimates of ellipticity existed,
tighter limits could be placed on halo triaxiality.

This paper describes our disk structure survey. With this survey, we
attempt to constrain the halo shapes of field galaxies by measuring disk
lopsidedness and ellipticity using a combination of bi-dimensional
spectroscopy and imaging, complementing other work such as the Fabry-Perot
kinematic survey of Virgo cluster disk galaxies (Chemin \etal 2006).
Beyond the implications such asymmetry
measurements have for cosmological simulations of galaxy halos, both
quantities also should be sources of scatter in the TF scaling
relation (Franx \& de Zeeuw 1992; Zaritsky \& Rix 1997; Andersen \&
Bershady 2003). Therefore, a measurement of ellipticity, and possibly
lopsidedness, places limits on other astrophysical sources of TF
scatter, including disk mass-to-light ratios. If limits are placed on
the variation of disk mass-to-light ratios, a new constraint will be
placed on mass decomposition models.

In this paper we present the H$\alpha$ and \HI~kinematic data for our
survey. H$\alpha$ echelle observations were made with the integral
field unit (IFU) DensePak (Barden, Sawyer \& Honeycutt 1998) on the
WIYN 3.5m telescope\footnote{The WIYN Observatory is a joint facility
of the University of Wisconsin-Madison, Indiana University, Yale
University, and the National Optical Astronomy Observatories.} for 39
nearly face-on disk galaxies. \HI~observations for 23 of the sample
come from the Nan\c{c}ay radio telescope. The optical data forms the
primary kinematic data for this survey. The purpose of the \HI~
line-profiles is to establish that the optical integral-field data
amply samples the potential. While \HI\ and H$\alpha$ have similar 
flux-weighted radial extents in many cases (Verheijen \& Sancisi 2001),
there will still exist different and varying spatial
distributions and ``turbulent'' motions of ionized and neutral gas
which could lead to significant differences between \HI~ and H$\alpha$
integrated line-widths. However, our expectations are that 
turbulent motions are small; the differences between ionized and
neutral turbulent motions smaller still; and the differences in
spatial distribution are mitigated by the typically flat, and steeply
rising rotation curves of this sample. Verifying these expectations is
important for our ultimate goal of measuring halo shapes, and also
allows us to address the suitability of optical line-widths as
dynamical tracers of total mass.  The application has relevance to
studies of both nearby and distant galaxies. The combination of
optical and radio data also is pertinent to understanding the origin
of line-profile asymmetries.

In future papers in this series we analyze the H$\alpha$ and \HI~
line-profiles, the H$\alpha$ velocity fields, and the photometric
properties of the survey galaxies. These studies explore the
photometric and kinematic asymmetries of the galaxy sample, and their
correlations. The studies build a physical picture of why asymmetries
arise; how they contribute to scatter in the Tully-Fisher relation;
and how well total dynamical mass can be measured in nearly face-on
systems. Together, they allows us to study the ellipticity
distribution of our sample in a tertiary analysis.

The present paper is organized as follows. We provide the selection
algorithm in \S 2. We present the DensePak and Nan\c{c}ay observations
in \S 3. The basic processing of the data is detailed in \S 4. \HI\
line-widths are also presented in \S 4. In \S 5 we describe how we combine
integral-field observations to obtain H$\alpha$ integrated
line-widths, present measurements of turbulent motions, discuss
integrated line-profiles of individual galaxies, and contrast optical
and radio line-profiles. We conclude with a summary in \S 6.

\section{Selection Algorithm}

The selection of the sample for this survey is driven by our desire to
measure disk ellipticity.  A 10\% disk ellipticity will create
differences in observable quantities which are smaller than typical
measurement errors for inclined galaxies (Barnes \& Sellwood 2003;
Andersen, Bershady, Sparke 2005), thereby necessitating that our
sample include only nearly face--on galaxies. To illustrate this,
Figure 1 shows how measurement errors on observable quantities
translate into intrinsic ellipticity errors as a function of
inclination. We have assumed that intrinsic ellipticity is measured
via a comparison of observed photometric and kinematic position
angles, photometric axis ratios, and kinematic inclinations. A
description of the method is found in Andersen \etal (2001). Given
the expected errors on these quantities and that the mean intrinsic
ellipticity of galaxies is expected to be below 10\%, galaxies with
inclinations $<$ 30$^\circ$are optimum choices for intrinsic
ellipticity measurements.

\begin{figure}
\vbox to 3.3in{\rule{0pt}{3.3in}}
\includegraphics{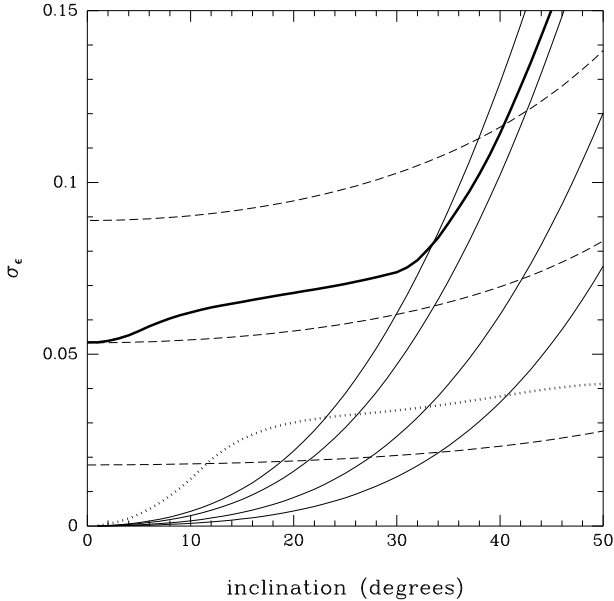}
   \caption{The 68\% confidence limit error distribution in intrinsic
disk ellipticity, $\sigma_\epsilon$, versus inclination. The lighter
solid lines represent the contribution to the ellipticity error due to
the measurement errors of 5$^\circ$, 7$^\circ$, 10$^\circ$ and
12$^\circ$ in the difference between kinematic and photometric
position angles ($\sigma_\psi$).  The dashed lines are the
contribution to the error on ellipticity due to 0.01, 0.03 and 0.05
axis ratio measurement errors ($\sigma_{b/a}$).  The dotted line is
the ellipticity error due to kinematic inclination errors
($\sigma_i$).  For each inclination angle, we used an empirical
relationship between inclination and kinematic inclination to estimate
the error (Andersen \& Bershady 2003). The heavy solid line is the
total ellipticity error as a function of inclination for the
inclination error, $\sigma_\psi =10^\circ$ and $\sigma_{b/a} =$0.03.
Ellipticity errors were calculated assuming an ellipticity of 0.1 and
that the intrinsic ellipticity major axis is aligned with the
kinematic major axis.  While the details of these curves vary slightly
for different ellipticities and phase angles between the ellipticity
and kinematic axes, it is always true that ellipticity errors increase
with inclination.  }\end{figure}

Two methods exist for making a selection of nearly face--on disks: 
Galaxies can be chosen with photometric axis ratios close to unity,
or chosen with narrow \HI~line-widths (or other kinematic measures of
rotation). The Tully-Fisher relationship can be inverted; line widths
and absolute magnitudes can used to solve for inclination (Rix \& Zaritsky
1995). Because the scatter in the Tully-Fisher relation is small for
normal spirals, these so-called ``inverse Tully-Fisher'' inclinations
could be used to select galaxies with inclinations less than
30$^\circ$. 

Choosing galaxies with existing \HI~line-widths may bias a sample.
Many galaxies are selected for \HI~observations on the basis of having
photometric inclinations greater than 45$^\circ$ for inclusion in
Tully-Fisher studies. Those galaxies with narrow \HI~line-widths may
indeed be nearly face--on, but in this case are necessarily highly
elliptic. Hence surveys built upon the availability of \HI~line-widths
as a selection criterion may preferentially include spiral galaxies
with elliptical disks.  We therefore choose galaxies for this study
based on photometric axis ratios.

However, disks with the greatest ellipticities will appear nearly
face--on only if the intrinsic minor axis lies near the inclination
axis. This potential bias is a second-order effect which can be
accounted for when modeling the distribution in ellipticities for the
general population of spiral galaxies (Ryden 2004). We therefore chose
galaxies with axis ratios near unity.

The DensePak integral field unit on the WIYN telescope spans 30 arcsec
$\times$ 45 arcsec, so targets were limited to diameters ($D_{25}$)
between 45 and 75 arcseconds so they could be well sampled in a few
pointings. Given this relatively small size, chosen galaxies will
typically be some of the faintest and most distant galaxies in the
catalogs of nearby galaxies.

To maximize the number of
galaxies available for study, we chose our sample from the LEDA
database\footnote{http://leda.univ-lyon1.fr} which contains all
galaxies in common catalogs (e.g. NGC, UGC, IC), plus additional
galaxies only defined in the Principal Galaxy Catalog (PGC; Paturel
\etal 1997).  While the PGC is inclusive of many other galaxy
catalogs, galaxies with diameters less than $\sim$1 arcminute still
fall below the completeness threshold (Figure 2).
We required sample galaxies from LEDA to have:
(1) axis ratio $b/a + \Delta b/a >
0.87$ ($i - \Delta i <30^\circ$)\footnote{Here, $\Delta b/a$ is the
  uncertainty in the measured axial ratio $b/a$, and $ \Delta i$ is
  the corresponding uncertainty in inclination as tabulated in LEDA.},
(2) apparent
disk size $45^{\prime\prime} < D_{25} < 75^{\prime\prime}$,
(3) galactic absorption $A_B < 0.6$
(Schlegel, Finkbeiner \& Davis 1998), (4) declinations visible from
WIYN, $\delta_{1950} > -10^\circ$, (5) observed $B$-band magnitudes,
(6) observed recession velocities, and (7) {\sl t}-types ranging from
Sab to Sd ({\sl t}-types between 1.5--8.5).  We elected not to observe
Sa galaxies because these galaxies have less H$\alpha$ emission than
later types (Kennicutt, Edgar \& Hodge 1989).  Of the 1300 targets
which met these raw selection criteria, 753 were observable during our
longest WIYN DensePak runs (January and December 1999; $0^h<RA<13^h$).

\begin{figure}
\vbox to 5.2in{\rule{0pt}{5.2in}}
\includegraphics{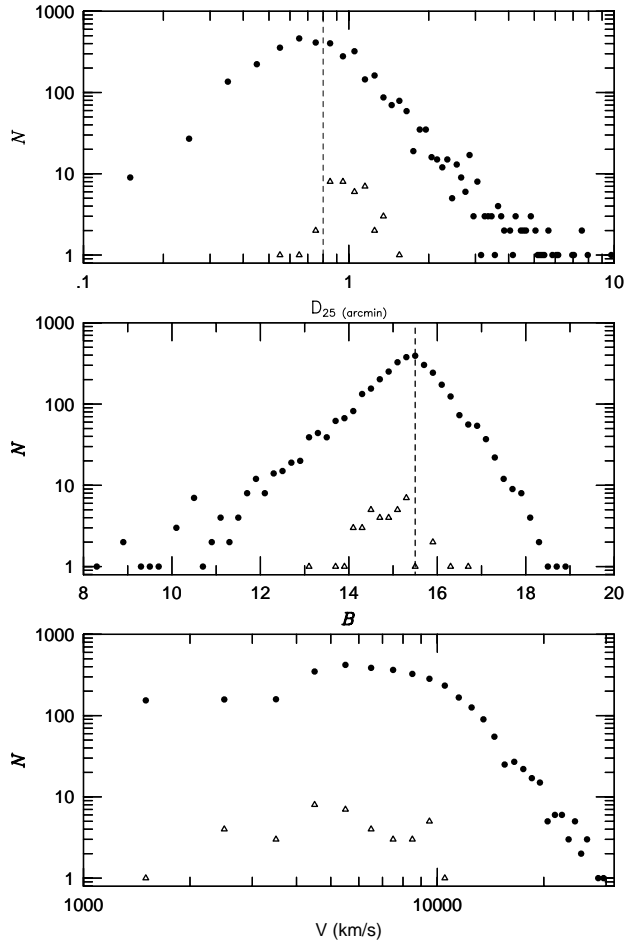}
\caption{Distribution of diameters (D$_{25}$; top panel), blue
magnitudes ($B$; middle panel) and recession velocities (V; bottom
panel) for galaxies in the PGC meeting the selection criteria in \S
2.2 (filled circles) and galaxies observed as part of this survey
(open triangles).  Specifically, these selection criteria include: (1)
$1.5 < t$-type $<8.5$, (2) $A_B<0.6$, (3) $\delta_{1950}>-10^\circ$,
(4) $b/a_{25} + \Delta b/a_{25} > 0.87$.  Completeness criteria for
galaxies in the PGC are met for D$_{25} > 0.8^\prime$ and $B<15.5$.
Despite the incompleteness of the parent sample, we endeavored to
choose a representative sample of disk galaxies which spanned a wide
range in types and luminosities.}
\end{figure}

We obtained Second Generation Digitized Sky Survey (DSS2)
images to further refine our sample selection.  After visual
inspection, we eliminated from our sample galaxies with bars, rings,
interacting companions, or foreground stars that contaminated the
inner isophotes. This step removed 60\% of the sample from
consideration. We measured scale lengths, position angles and axis ratios
from the
DSS2 images using the {\sl IRAF} {\it ellipse} routine, and we further 
required (1) a relatively constant position angle in the outer
portions of the disk, therein removing most warped galaxies, and (2)
our measurements of $b/a>0.87$ on DSS2 images at three scale lengths.
Between $0^h < RA < 13^h$ only 70 targets remained. As a test
of the completeness of our selection criteria, we examined the
DSS2 images of every PGC galaxy in a 1 hour slice of right ascension
and ascertained that our selection criteria successfully yielded every
galaxy with a photometric inclination less than 30$^\circ$ in this
slice of right ascension.  Through the course of our observations, we
observed 32 of these 70 targets, plus 7 other galaxies chosen from the
same parent sample of 1300 targets, but which lie between $13^h < RA <
24^h$.  The 39 observed galaxies are not representative of the parent
sample, rather we attempted to span a large range of morphological types and
luminosity (Table 1 and Figure 3).

\begin{deluxetable*}{llllllllllll}
\tabletypesize{\tiny}
\tablewidth{0pt} 
\tablecaption{Properties of survey galaxies tabulated from the 
PGC\tablenotemark{a}}
\tablenum{1}
\tablehead{
\multicolumn{1}{l}{PGC}   &
\multicolumn{1}{l}{alternate}   &
\multicolumn{1}{l}{RA}   &
\multicolumn{1}{l}{DEC}   &
\multicolumn{1}{l}{type\tablenotemark{b}}   &
\multicolumn{1}{l}{t-type\tablenotemark{c}} &
\multicolumn{1}{l}{D$_{25}$\tablenotemark{d}} &
\multicolumn{1}{l}{$b/a_{25}$\tablenotemark{e}}  &
\multicolumn{1}{l}{m$_B$}   &
\multicolumn{1}{l}{A$_B$\tablenotemark{f}}   &
\multicolumn{1}{l}{V$_{\rm helio}$}  &
\multicolumn{1}{l}{M$_B$\tablenotemark{g}} \\
\multicolumn{1}{l}{} &
\multicolumn{1}{l}{name} &
\multicolumn{1}{l}{(1950)} &
\multicolumn{1}{l}{(1950)} &
\multicolumn{1}{l}{} &
\multicolumn{1}{l}{} &
\multicolumn{1}{l}{(arcsec)} &
\multicolumn{1}{l}{} &
\multicolumn{1}{l}{(mag)} &
\multicolumn{1}{l}{(mag)} &
\multicolumn{1}{l}{(km s$^{-1}$)} &
\multicolumn{1}{l}{(mag)}
}
\startdata
 2162 & UGC 358    & 00 33 39.0 & +01 26 13 & Sab  & 1.8 & 0.8 & 0.90 & 15.5 & 0.08 & 5448 & -18.9 \\
 3512 & IC  1607   & 00 56 15.0 & +00 19 04 & Sb   & 3.0 & 0.9 & 0.87 & 14.4 & 0.14 & 5435 & -20.0 \\
 5345 & UGC 1014   & 01 23 47.1 & +06 01 04 & SBm  & 9.2 & 1.1 & 0.80 & 15.2 & 0.14 & 2132 & -17.3 \\
 5673 & UGC 1087   & 01 28 46.5 & +14 01 13 & Sc   & 5.3 & 1.2 & 0.93 & 14.8 & 0.23 & 4485 & -19.3 \\
 6855 & UGC 1322   & 01 48 49.1 & +12 53 04 & Sc   & 5.2 & 1.1 & 0.81 & 15.8 & 0.36 & 4834 & -18.7 \\
 7826 & UGC 1546   & 02 00 35.1 & +18 23 22 & Sc   & 5.4 & 0.9 & 0.92 & 15.2 & 0.32 & 2371 & -17.6 \\
 8941 & UGC 1808   & 02 18 16.0 & +23 22 20 & Sb   & 3.1 & 0.9 & 0.92 & 14.9 & 0.39 & 9447 & -21.0 \\
14564 & NGC 1517   & 04 06 29.2 & +08 31 04 & Sc   & 6.0 & 1.0 & 0.85 & 14.2 & 1.10 & 3483 & -20.2 \\
15531 & UGC 3091   & 04 31 21.2 & +01 00 36 & Scd  & 6.7 & 1.1 & 0.85 & 15.1 & 0.36 & 5557 & -19.6 \\
16274 & MCG 1-13-8 & 04 51 19.3 & +03 30 46 & Sb   & 2.7 & 0.8 & 1.00 & 15.0 & 0.32 & 8915 & -20.6 \\
19767 & UGC 3569   & 06 48 12.6 & +57 13 23 & Sd   & 7.6 & 0.9 & 0.83 & 16.3 & 0.22 & 5133 & -18.2 \\
20938 & UGC 3833   & 07 21 33.4 & +32 54 12 & Sbc  & 4.3 & 0.5 & 0.88 & 15.9 & 0.22 & 4695 & -18.3 \\
23333 & IC  2283   & 08 16 16.2 & +24 57 00 & Sb   & 3.3 & 0.9 & 0.86 & 14.6 & 0.14 & 4654 & -19.5 \\
23598 & UGC 4380   & 08 20 40.7 & +55 00 58 & Sc   & 5.9 & 0.9 & 0.95 & 15.3 & 0.26 & 7485 & -20.0 \\
23913 & UGC 4445   & 08 27 17.2 & +61 09 50 & Sc   & 5.3 & 1.1 & 0.97 & 15.3 & 0.30 & 6330 & -19.7 \\
24788 & UGC 4614   & 08 46 06.2 & +36 18 20 & SBb  & 2.9 & 0.7 & 0.83 & 15.1 & 0.15 & 7556 & -20.2 \\
26140 & NGC 2794   & 09 13 14.0 & +17 47 54 & SBbc & 4.4 & 1.1 & 0.83 & 14.0 & 0.14 & 8760 & -21.6 \\
26517 & UGC 4978   & 09 19 29.5 & +04 06 39 & Scd  & 6.6 & 1.5 & 0.84 & 15.2 & 0.20 & 4135 & -18.7 \\
27792 & UGC 5187   & 09 39 59.1 & +41 19 19 & SBbc & 4.3 & 0.9 & 0.81 & 14.7 & 0.08 & 1465 & -17.1 \\
28310 & UGC 5274   & 09 47 27.1 & +16 31 12 & SBc  & 5.9 & 1.1 & 0.99 & 14.7 & 0.18 & 5908 & -19.9 \\
28401 & UGC 5277   & 09 47 46.8 & +65 43 32 & SBbc & 3.6 & 1.3 & 0.90 & 14.4 & 0.71 & 3365 & -19.7 \\
31159 & IC  616    & 10 30 06.6 & +16 07 06 & Sc   & 5.9 & 1.0 & 0.85 & 14.8 & 0.17 & 5779 & -19.9 \\
32091 & MCG -2-28-6& 10 42 40.9 & -09 48 04 & Scd  & 6.9 & 1.3 & 0.92 & 14.2 & 0.15 & 2511 & -18.4 \\
32638 & NGC 3438   & 10 49 48.6 & +10 48 45 & Sbc  & 3.8 & 0.8 & 0.93 & 14.1 & 0.10 & 6488 & -20.7 \\
33465 & UGC 6135   & 11 01 46.7 & +45 23 41 & Sbc  & 3.7 & 0.9 & 0.95 & 13.1 & 0.03 & 5948 & -21.5 \\
36925 & NGC 3890   & 11 46 33.4 & +74 34 49 & Sc   & 4.8 & 0.8 & 0.92 & 14.4 & 0.26 & 6827 & -20.7 \\
38268 & UGC 7072   & 12 02 39.6 & +29 03 37 & Sd   & 8.0 & 1.2 & 0.85 & 15.3 & 0.08 & 3152 & -18.0 \\
38908 & UGC 7208   & 12 09 56.8 & +39 23 20 & Sbc  & 4.4 & 1.0 & 0.91 & 15.0 & 0.13 & 7078 & -20.1 \\
39728 & NGC 4275   & 12 17 21.8 & +27 53 54 & Sb   & 2.8 & 0.8 & 0.90 & 14.0 & 0.09 & 2317 & -18.6 \\
46767 & NGC 5123   & 13 20 58.7 & +43 20 50 & Sc   & 5.9 & 1.1 & 0.89 & 13.7 & 0.06 & 8323 & -21.6 \\
49906 & NGC 5405   & 13 58 40.3 & +07 56 35 & Sc   & 4.8 & 0.8 & 0.94 & 14.9 & 0.17 & 6922 & -20.0 \\
55750 & IC  132    & 15 37 53.7 & +20 50 28 & Sc   & 5.4 & 1.0 & 0.92 & 14.5 & 0.27 & 4525 & -19.8 \\
56010 & MCG 6-35-5 & 15 44 37.1 & +33 22 38 & SBc  & 6.2 & 0.8 & 0.96 & 15.3 & 0.14 & 4468 & -18.8 \\
57931 & UGC 10357  & 16 20 18.3 & +40 33 50 & Sbc  & 3.5 & 0.8 & 0.88 & 15.1 & 0.05 & 9280 & -20.5 \\
58410 & UGC 10436  & 16 29 24.2 & +41 15 44 & SBc  & 5.3 & 1.0 & 0.97 & 14.5 & 0.05 & 9059 & -21.0 \\
70962 & MRK 318    & 23 15 05.9 & +13 43 38 & Sbc  & 3.8 & 0.7 & 0.80 & 14.2 & 0.20 & 4455 & -20.0 \\
71106 & NGC 7620   & 23 17 37.3 & +23 56 49 & Sc   & 5.9 & 1.0 & 0.96 & 13.8 & 0.34 & 9582 & -22.1 \\
72144 & UGC 12740  & 23 39 22.1 & +23 32 15 & Sc   & 6.0 & 0.6 & 0.95 & 16.6 & 0.27 & 10521& -19.4 \\
72453 & UGC 12784  & 23 45 32.0 & +17 11 49 & SBbc & 3.6 & 1.3 & 1.00 & 14.7 & 0.30 & 9952 & -21.2
\enddata
\tablenotetext{a}{Paturel \etal (1997)}
\tablenotetext{b}{de Vaucouleurs type for the galaxy (de Vaucouleurs
\etal 1991).}
\tablenotetext{c}{Morphological type placed in a pseudo-numerical
scale (Paturel \etal 1997).}
\tablenotetext{d}{Diameter measured from the isophote at a $B$-band
surface brightness of $\mu_B=25$ mag arcsec${-2}$.}
\tablenotetext{e}{Axis ratio measured from the isophote at a $B$-band
surface brightness of $\mu_B=25$ mag arcsec${-2}$.}
\tablenotetext{f}{A$_B$ is calculated using the galactic absorption
law of Cardelli \etal (1989).}
\tablenotetext{g}{Absolute $B-$band magnitude adopts $H_0=75$ km
s$^{-1}$ Mpc$^{-1}$ and heliocentric recession velocity corrected for
peculiar motion associated with the local group and Virgo cluster.}
\end{deluxetable*}

\begin{figure}
\vbox to 5.2in{\rule{0pt}{5.2in}}
\includegraphics{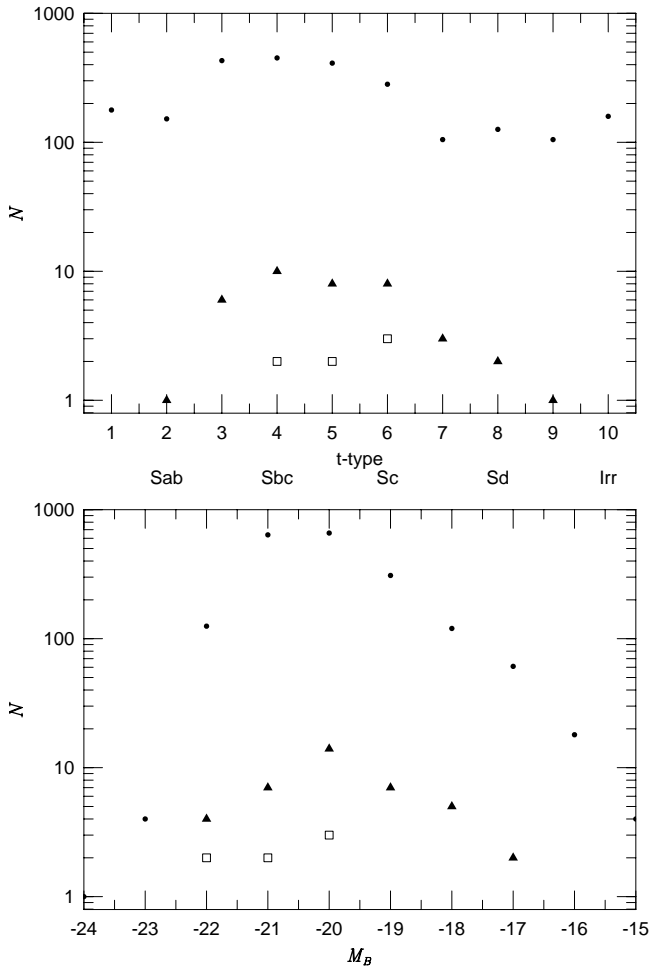}
\caption{The distribution in type (top panel) and luminosity (bottom
panel) for galaxies in the PGC meeting the selection criteria in \S
2.2 (filled circles) and galaxies observed as part of this study
(filled triangles).  The selection criteria applied to sample of
galaxies in the top panel include, (1) measured $B$-magnitudes and (2)
recession velocities, (3) $A_B<0.6$, (4) $\delta_{1950}>-10^\circ$,
(5) $b/a_{25} + \Delta b/a_{25} > 0.87$, and (6) $0.^\prime6
<D_{25}<2^\prime$.  The bottom panel include all galaxies in the top
panel which have $1.5 < t$-type $<8.5$.  We broadened the sample by
both type and luminosity upon completion of our pilot study (open
squares; Andersen \etal 2001).}
\end{figure}

\section{Kinematic Data}

\subsection{DensePak H$\alpha$ Observations}

We observed 39 sample galaxies with DensePak during 11 nights on 6
separate observing runs (Table 2).  The DensePak IFU is a fiber
optic--array mounted at the Nasmyth f/6.3 focus imaging port on the
WIYN 3.5m telescope. The WIYN Observatory, located on Kitt Peak south
of Tucson, Arizona, stands 6875 feet above sea level and is home to
the second largest telescope on the mountain.  Dedicated in October,
1994, the WIYN telescope employs an altitude--azimuth mount design.
DensePak is an array of 91 fibers arranged in a seven by thirteen
fiber rectangle sub-tending an area of 28$^{\prime\prime}\times$
42$^{\prime\prime}$.  Fibers have an active core diameter of
2.81$^{\prime\prime}$ (300 $\mu$m).  Cladding and buffer increase the
total fiber diameter to 3.75$^{\prime\prime}$ (400 $\mu$m). According
to the DensePak manual's astrometric diagram (Sawyer 1997, Figure 4)
\footnote{This manual is currently available at
www.noao.edu/wiyn/densepak.pdf. The astrometric diagram was
subsequently published in Homeier \& Gallagher (1999). Note there is
some uncertainty in the adopted fiber spacing.  The plate scale of 10
arcsec mm$^{-1}$ quote in the manual is incorrect, and should be 9.374
arcsec mm$^{-1}$ for the bare RC focus of the WIYN port where DensePak
was used (see, for example: www.noao.edu/wiyn/wiynfacts.html). This
reconciles the different fiber diameters quoted here versus the manual
and Homeier \& Gallagher. However, it is unclear if the manual's
astrometric diagram is based on this erroneous plate scale. For the
purpose of this work we adopt the implied fiber spacing given by the
numbers in the astrometric diagram.} the fiber spacing is 4.2 arcsec
(center-to-center).  In addition to the 91 fibers arranged in a
rectangle, another 4 fibers are spaced around the rectangle roughly an
arcminute from the center and are used to measure the ``sky'' flux. Of
the original 91 DensePak fibers, 86 fibers remain --- 5 fibers are
broken (Figure 4).

\begin{deluxetable}{lcllc}
\tabletypesize{\tiny}
\tablewidth{0pt}
\tablecaption{DensePak Observing Log}
\tablenum{2}
\tablehead{
\multicolumn{1}{l}{PGC} &
\multicolumn{1}{c}{Run ID\tablenotemark{a}}&
\multicolumn{1}{l}{U.T. date} &
\multicolumn{1}{l}{Exposure Time} &
\multicolumn{1}{c}{pointings} \\
\colhead{} &
\colhead{} &
\colhead{} &
\multicolumn{1}{l}{(sec)} &
\colhead{}
}
\startdata
02162 & e & 12/22/99 & 3600, 3600 & 2 \\
03512 & e & 12/21/99 & 3600, 2400 & 2 \\
05345 & e & 12/20/99 & 3600, 3600 & 2 \\
05673 & f & 12/29/00 & 3600, 3600, 3000 & 3 \\
06855 & f & 12/30/00 & 3000, 3000, 3000 & 3 \\
07826 & e & 12/22/99 & 3600, 3600 & 2 \\ 
08941 & b & 01/23/99 & 3600, 3600 & 2 \\
14564 & b & 01/22/99 & 3600, 1800 & 2 \\
15531 & b & 01/23/99 & 3600, 3600, 3600 & 3 \\
16274 & e & 12/19/99 & 3600, 3600 & 2 \\
19767 & f & 12/29/00 & 3000, 3000 & 2 \\
20938 & e & 12/19/99 & 3600, 3600, 2400, 2400 & 4 \\
23333 & e & 12/22/99 & 3600, 3600 & 2 \\
23598 & b & 01/22/99 & 3600, 3600 & 2 \\
23913 & b & 01/22/99 & 3600, 3600 & 2 \\
24788 & f & 12/30/00 & 2400, 2400 & 2 \\
26140 & b & 01/21/99 & 3600 & 1 \\
26140 & f & 12/29/00 & 2400, 2400 & 2 \\
26517 & f & 12/30/00 & 3000, 3000 & 2 \\
27792 & e & 12/21/99 & 3600, 3600 & 2 \\
28310 & b & 01/22/99 & 3600, 3600 & 2 \\
28401 & e & 12/19/99 & 2400, 1800 & 2 \\
31159 & c & 03/28/99 & 3600 & 1 \\
31159 & f & 12/29/00 & 3000, 2400 & 2 \\
32091 & e & 12/21/99 & 3600, 3600 & 2 \\
32638 & f & 12/30/00 & 2400, 2400 & 2 \\ 
33465 & e & 12/22/99 & 2400, 2400 & 2 \\
36925 & c & 03/28/99 & 3600 & 1 \\
38268 & e & 12/21/99 & 3600, 3600 & 2 \\
38908 & b & 01/22/99 & 3600 & 1 \\
38908 & f & 12/30/00 & 3000, 3000 & 2 \\
39728 & e & 12/22/99 & 2400 & 1 \\
46767 & b & 01/22/99 & 3600, 3600 & 2 \\
46767 & b & 01/23/99 & 3600 & 1 \\
49906 & c & 03/28/99 & 3600 & 1 \\
49906 & f & 12/29/00 & 3000, 3000 & 2 \\
55750 & c & 03/28/99 & 3600, 3000 & 2 \\
56010 & c & 03/28/99 & 3600, 1200 & 2 \\
56010 & c & 03/29/99 & 2400 & 1 \\
57931 & d & 09/03/99 & 3600, 3600 & 2 \\
58410 & a & 05/22/98 & 3600, 3600 & 2 \\
70962 & e & 12/22/99 & 2400 & 1 \\
71106 & e & 12/20/99 & 2400, 2400 & 2 \\
72144 & e & 12/19/99 & 3600, 3600 & 2 \\
72453 & e & 12/21/99 & 3600, 3600 & 2
\enddata
\tablenotetext{a}{The Run Identification identifies the Principal
Investigator (PI) and observers for each run:
(a) PI: Bershady (UW allocation), Observers: Bershady \& Andersen;
(b) PI: Bershady, Gallagher, Sparke, Wilcots (UW allocation), Observer:
Andersen;
(c) PI: Bershady, Gallagher, Sparke, Wilcots (UW allocation), Observers:
Andersen \& G. Madsen;
(d) PI: Bershady (UW allocation), Observer: Andersen;
(e) PI: Andersen (NOAO allocation), Observer: Andersen;
(f) PI: Bershady \& Andersen (UW allocation), Observer: Andersen.}
\end{deluxetable}

\begin{figure}
\vbox to 3.4in{\rule{0pt}{3.4in}}
\includegraphics{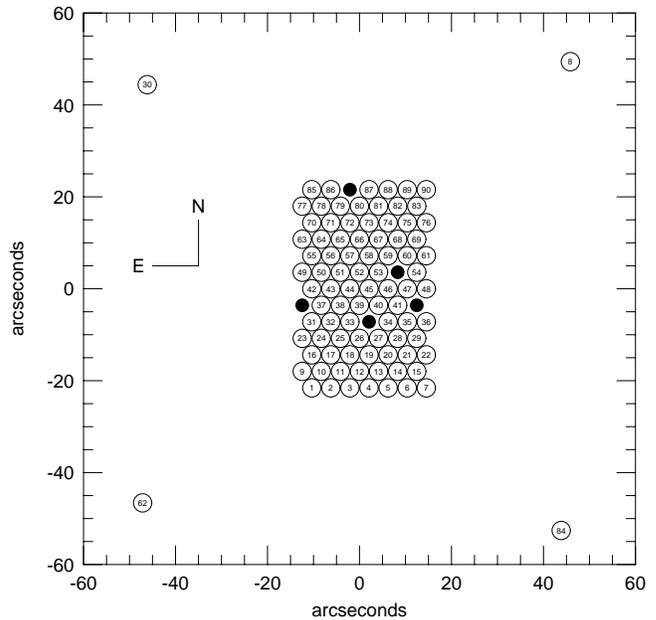}
\caption{Schematic of DensePak in the telescope focal plane.  Fibers
are identified according to their position in the output slit. There
are empty spaces in the slit between fibers 22--23, 26--27, 33--34,
36--37, 41--42, 53--54 and 86--87. Solid black disks mark broken
fibers in the fiber bundle. Spatial orientation is for a zero rotation
offset. Fibers are numbered in order of live fibers along the
reconfigured output slit. This numbering scheme differs slightly from
the DensePak manual.} 
\end{figure}

DensePak feeds the WIYN Bench Spectrograph, a fiber--fed spectrograph
designed to provide low to medium resolution spectra.  We used the
Bench Spectrograph camera (BSC) and 316 lines/mm echelle grating in
order 8 to cover 6500\AA $< \lambda\lambda < 6900$\AA, with a
dispersion of 0.195 \AA/pix (8.6 km s$^{-1}$/pix) and an instrumental FWHM of
0.51 \AA~(22.5 km s$^{-1}$).  We required the highest resolution available
with this instrument; fitting velocity field models to galaxies with
observed rotation velocities of less than 100 km s$^{-1}$ required centroids
with accuracies of roughly 5 km s$^{-1}$.  The BSC images the spectrograph
onto a T2KC thinned SITe 2048x2048 CCD with 24 $\mu$m-pixels. The
spectra are aligned along the columns of the CCD. The chip has a read
noise of 4.3 e$^-$ and was used with the standard gain of 1.7
e$^-$/ADU.  The system throughput for this setup is roughly 4\%
estimated from comparing our mean spectral continuum fluxes to
calibrated $R$-band images (Bershady \etal 2005).

Since galaxy rotation curves typically peak at roughly two scale
lengths (Courteau \& Rix 1999; Willick 1999) and H$\alpha$ emission is
only detected out to 3--4 scale lengths (Rubin, Waterman \& Kenney
1999; Dale \etal 2001), our goal was to spatially cover galaxies out
to 2.5--3.5 scale lengths.  
Galaxies with DSS2 scale
lengths less than six arcseconds only required one DensePak position
to cover several scale lengths. However, to improve fits to the
velocity field we required that two spatial resolution elements (fibers)
be used to sample one disk scale length. For galaxies with scale lengths
less than six arcseconds, two
overlapping DensePak positions offset by a half-fiber diameter were
used.  For scale lengths between 6" and 9", we used two DensePak
positions offset to cover a 45"$\times$45" area, and for scale lengths
between 9" and 12", we used three DensePak positions. The first two
positions were offset 27" from each other.  For the third position,
DensePak was rotated, centered, and offset south of the first two
positions. These three DensePak positions covered almost one square
arcminute.  In order to observe the H$\alpha$ line emission at three
scale lengths and improve cosmic-ray rejection, we typically took two,
$\sim$30 minute exposures at each position.

To center DensePak on our target we used a slit-viewing camera built
into the fiber-mounting module known as WIFOE. Use of this camera
involves inserting a pellicle for simultaneous viewing of the source
and the rear-illuminated IFU. In this way, we aligned fiber 45 on a
nearby guide star, and then zeroed the telescope offsets.  After
moving to our target, we could see the target galaxy in the
slit-viewing camera only if the galaxy had a high central surface
brightness and the sky was dark and clear. If we saw the galaxy, we
confirmed it was centered on fiber 45 and performed the first
offset. If the galaxy was not visible, we performed a blind
offset. After guiding was established, we began
observations. Typically, after two thirty minute exposures, guiding
was turned off and the second set of telescope offsets were applied.
After guiding was re-established, we took our next two exposures.  If
a third position was required, we re-acquired the nearby guide star
and rotated DensePak. Since DensePak was not on the rotator axis, the
star shifted a few arcseconds as DensePak was rotated 90$^\circ$.
After re-centering the guide star on fiber 45, we moved DensePak back
to the source, applied the final telescope offset and took our final
two exposures after re-establishing guiding. We found that rotating
DensePak added up to ten minutes of overhead to our observations; we
therefore preferentially chose galaxies which required only two
DensePak footprints to sample the disk out to 2.5 scale lengths.

We encountered problems with accurately positioning DensePak on-sky
due to problems with the pellicle and non-guided offsets. For some of
our observations, the pellicle did not properly align images in the
slit-viewing camera, so our spectra were not centered as
expected. However, when fitting velocity field models to multiple
pointings, model centers were easily determined from the data. Hence,
the unaligned pellicle did not diminish the quality of the
data. However, inaccurate telescope offsets did complicate data
handling.  We performed tests at the telescope on the accuracy of our
offsets by removing each offset in turn and re-centering on the guide
star. We discovered that the guide star moved as much as a fiber
diameter (3 arcseconds) after removing the telescope offsets.
Therefore, we are forced to let telescope offsets be a free parameter
in our model fits, thereby introducing extra fit parameters which
could have been avoided if reliable offsets were possible. More tests
are needed to determine an accurate method for applying telescope
offsets using DensePak on WIYN.

At the beginning and end of each night, dome flats and bias frames
were taken. We also observed Thorium Argon (ThAr) line lamps at the
beginning of the night, before and after the dewar was filled, and
again at the end of the night. The BSC dewar has a short hold time of
only 12 hours, and typically was refilled during the course of the
night. We observed sub-pixel shifts of $\sim 0.1$ pixels in the
centroids of the ThAr emission-lines taken at different times. The
largest shifts are seen between frames taken before and after dewar
fills. When reducing spectra, we calibrated wavelengths using the ThAr
frame taken closest to a given observation, within a period between
filling of the dewar.

The quality of data from DensePak varied across the slit formed by the
90 fibers. In particular, comparisons of dome flat fluxes showed the
first eight fibers in the top of the slit had less than half the flux
of most fibers (see Figure 3 of Bershady \etal 2005). After field
flattening, this low throughput was translated into greater values for
the continuum noise in these fibers. We believe this poor performance
may be due to: (1) fibers 1-8 undergo the tightest bend near the slit
end of DensePak of all the fibers.  Bershady \etal (2004) present
evidence that this tight bend radius increases the focal ratio
degradation of fibers for a similar integral field unit, SparsePak,
thereby leading to greater losses in the spectrograph; (2) fibers 1-8
are at the top end of the DensePak slit. If the slit is not centered
within the baffles at the spectrograph entrance, these fibers could
suffer from further vignetting; (3) alternatively, these fibers may be
otherwise over-stressed, e.g. because these fibers are on an outer
face of the array (see Bershady \etal 2004).
Data from fiber 1 were so poor that we did not include them when, for 
example, we stacked spectra to make synthetic line profiles (\S 5.2).

\subsection{Nan\c{c}ay \HI~Observations}

The Nan\c{c}ay telescope is a meridian transit-type instrument of the
Kraus/Ohio State design, consisting of a fixed spherical mirror, 300 m
long and 35 m high, a tiltable flat mirror (200$\times$40 m), and a
focal carriage moving along a 90 m long curved rail track, which
allows the tracking of a source on the celestial equator for about 1
hour. Located in the center of France, it can reach declinations as
low as -39$^\circ$. It has an effective collecting area of roughly
7000~m$^{2}$ (equivalent to a 94-m diameter parabolic dish).  Due to
the elongated geometry of the mirrors, at 21-cm wavelength it has a
half-power beam width (HPBW) of \am{3}{6}~E-W$\times$ 22$'$~N-S for
declinations below 30$^{\circ}$; at higher declinations the N-S HPBW
increases (see plot in Matthews, van Driel \& Monnier-Ragaigne 2001).
Typical system temperatures were $\sim$40~K for our project. For a
technical description of the Nan\c{c}ay decimetric radio telescope and
the general methods for data handling and reduction see Theureau et
al. (1998a) and references therein.

Observations at Nan\c{c}ay of 23 sample galaxies were made in the
periods March to August 1999 and January to June 2001 using a total of
about 110 hours of telescope time.  We obtained our observations in
total power (position-switching) mode using consecutive pairs of
two-minute on- and two-minute off-source integrations. Off-source
integrations were taken at approximately 20$'$~E of the target
position.  The autocorrelator was divided into two pairs of
cross-polarized receiver banks, each with 512 channels and a 6.4~MHz
bandpass. This yielded a channel spacing of 2.64~km s$^{-1}$, for an
effective velocity resolution of $\sim$3.3~km s$^{-1}$ at 21-cm, which was
smoothed to a channel separation of 7.91 and a velocity resolution of
9.50 km s$^{-1}$ during the data reduction, in order to search for faint
features. The center frequencies of the two banks were tuned to the
known redshifted \HI~frequency of the target.

\section{Basic Processing}

\subsection{DensePak Reductions}

Data were overscan and bias-corrected and trimmed using the NOAO {\sl
IRAF} package {\it ccdproc}. Since two or more frames were taken at each
position, we calculated pair-wise differences and statistical
thresholds, pixel-by-pixel, of the expected variance ($\sigma^2$) due
to read-noise and shot-noise. Pixels which were more than 5$\sigma$
brighter than the corresponding pixels on the other frames were flagged
as cosmic rays and masked from the final, mean-combined, two-dimensional
spectral image. This is preferable to removing cosmic rays
after spectral extraction which averages cosmic rays falling in pixels
that are part of extracted apertures with good data from other pixels
in same spectral channel. Following cosmic-ray cleaning, basic
spectral extraction,
flattening, wavelength calibration and sky subtraction were done using
{\sl IRAF} {\it dohydra}. 
Extracted one-dimensional spectra were
field-flattened with dome flats and wavelength-calibrated using ThAr
emission spectra.  Finally, the four sky spectra were averaged and the
mean sky spectrum was subtracted from the other 86 spectra.

Once spectra were processed with IRAF {\it dohydra}, we identified
extracted H$\alpha$ emission-line characteristics. 
Our algorithm measured Gaussian fluxes, widths, centers and centroid
errors for lines in a given spectral interval. Most
emission-lines in our data were symmetric and were well-fit by a
single Gaussian. Some H$\alpha$ profiles sampled by fibers within a
fiber radius of the galaxy center were skewed or even bimodal as would
be expected when the fiber diameter is larger than the local dynamical
scale.  Since the fraction of such lines was small, we flagged these
lines during visual inspection after running our automated line
fitting algorithm and do not cite the fit results later.
While Beauvais \& Bothun (1999) 
found the best fits to a variety of
different simulated line-profiles were obtained with Vogt profiles (a
combination of Gaussian and Lorentzian profiles), they noted
that Gaussian fits yielded comparable results for signal-to-noise (S/N)
ratios greater than 20. A similar conclusion was reached by Courteau (1997)
who measured intensity-weighted centroids.
Our typical H$\alpha$
lines had high S/N, and a visual inspection of the fits 
made it clear that Gaussian profiles were sufficient to 
describe most emission lines.

Since the S/N of each channel in the
spectra is important for accurate profile fitting, we iteratively
established measurement errors for each channel.  We begin by
calculating an iterative clipped mean and standard
deviation within a 30\AA~ window containing the emission line.
We used a very tight clipping criterion: All channels with
counts greater than the standard deviation were rejected, repeating
until the number of channels did not decrease further.  This procedure
removed almost all effects of emission-lines, imperfectly subtracted
sky lines, and residual cosmic rays from the continuum mean and
standard deviation.  We normalized the standard deviation of the
iteratively clipped data by the standard deviation of a Gaussian
distribution which also had been truncated at 1$\sigma$.  With the
continuum standard deviation, $\sigma_{\rm cont.}$, established, we
assumed remaining errors on channel counts were based on Poisson
counting errors. Therefore the error on each channel, $\sigma_i$, 
was calculated as follows: \begin{equation} \sigma_i =
\{\sigma_{\rm cont.}^2 + ({\rm Counts}_i - {\rm Counts}_{\rm
cont.})\}^{1/2} \end{equation} where all counts were calculated in
$e^-$/pixel, ${\rm Counts}_i$ was the counts for channel $i$, and
${\rm Counts}_{\rm cont.}$ was the iterative mean calculated in the
200\AA~ window.
 
Once electron counts and the initial estimates of the
standard deviations had been calculated for
every channel, we used the Marquardt--Levinson algorithm to fit a
Gaussian profile plus a linearly varying background to the line within
the selected region.  We then subtracted this best fit spectrum
within the window and recalculated the standard deviation following
the same procedure described above. Then we refit the data and extract the
Gaussian amplitude, center ($\lambda_C$) and width ($\sigma_G$) 
from the fit. The area beneath the
curve yields the flux ($f_G$).  The error on the centroid
($\sigma_{\lambda_C}$) was obtained from the covariance matrix.
 
\begin{figure*}
\vbox to 8.2in{\rule{0pt}{8.2in}}
\includegraphics{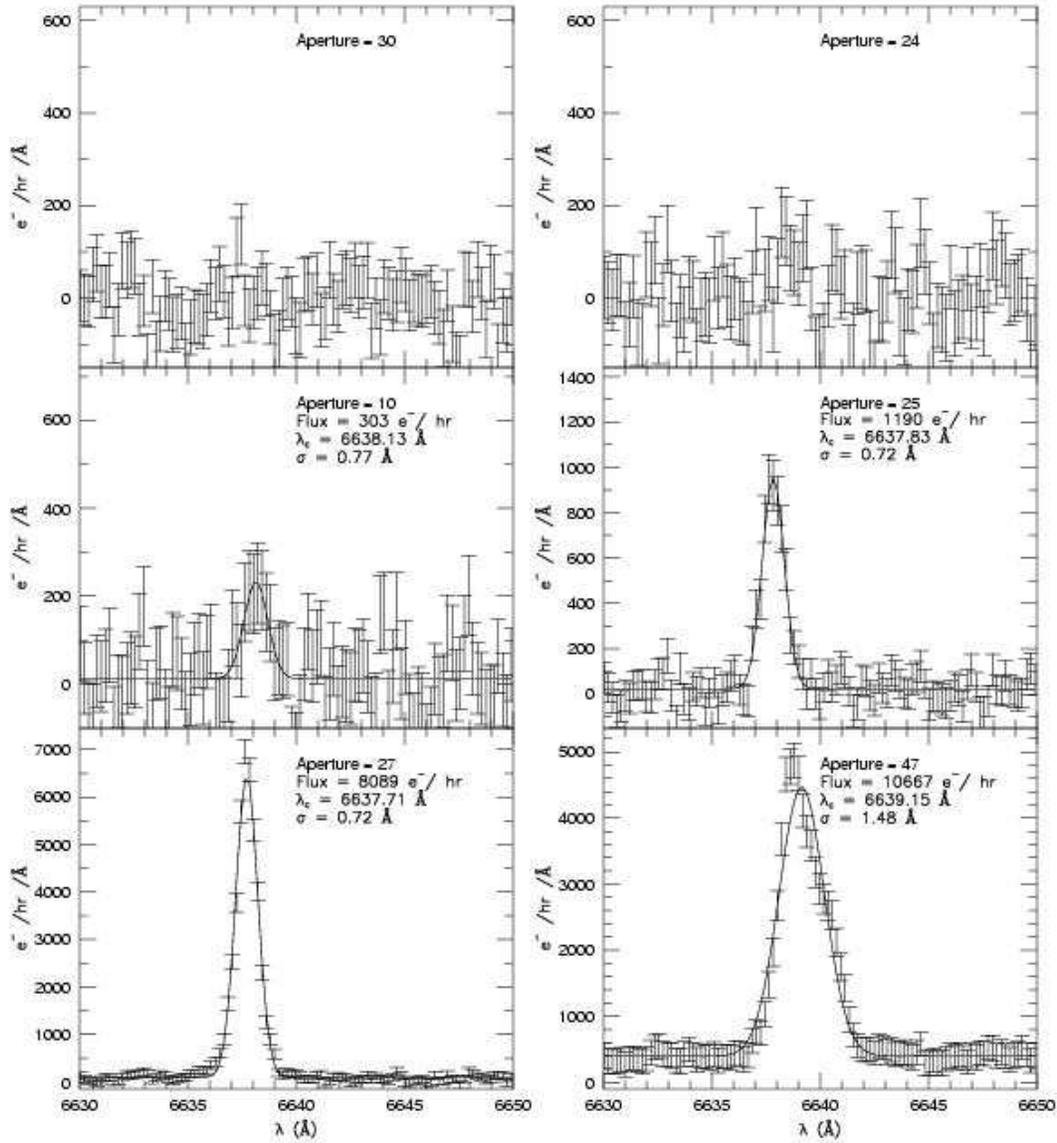}
\caption{H$\alpha$ emission-line-profiles for 6 fibers selected from
the first pointing of PGC 14564. Fiber 30 was a sky fiber and
illustrates the typical background fluctuations. The solution for a
Gaussian H$\alpha$ emission-line in Fiber 24 (upper right panel) did
not meet the S/N criterion: $\zeta\equiv f_G/\sigma_{\rm
cont.}/(2\sigma_G)^{1/2}>5$.  For the emission-line in aperture 10
(left, middle panel), $\zeta=8$ .  The H$\alpha$ emission-line for
aperture 25 (middle right panel) was typical for most observed
galaxies. Aperture 27 (bottom left) shows very strong H$\alpha$
emission for our survey. The H$\alpha$ profile in aperture 47 (lower
right panel) was highly skewed, showing signs of bimodality.}
\end{figure*}

We found that features which had Gaussian fits that did not reach the
S/N threshold $f_G/\sigma_{\rm cont.}/(2\sigma_G)^{1/2}>5$
were also not identified as emission-lines by visual inspection. We
also required the width of the line must be equal to or greater than
the resolution element. Finally, as already noted, we discarded lines
that, based on visual inspection, were skewed, required multiple
Gaussian profiles, or formally met the S/N and width requirements but
none-the-less did not look plausible.  Figure 5 shows a sample of the
output with Gaussian fits chosen for their range of properties: an
example of sky continuum, a line which did not meet the S/N threshold,
a weak line which did have sufficient S/N, typical, strong and skewed
lines.

Table 3 contains a sample of the DensePak line-fits for the first
pointing of PGC 14564. The line-fits tables for all 39 galaxies are
available in electronic form. Tabulated offsets refer to the fiber
positions West and North of the kinematic center based on our model
fits to the velocity fields described in the relevant forthcoming
paper in this series, and the astrometric information regarding
relative fiber positions described in \S 3.1.

\begin{longtable*}{ccccrrrrc}
\tabletypesize{\tiny}
\tablenum{3}
\tablewidth{0pt}
\tablecaption{Gaussian Fits to H$\alpha$ emission-lines: 1st pointing of PGC 14564}
\tablehead{
\colhead{Fiber} &
\colhead{Pointing} &
\colhead{West} &
\colhead{North} &
\colhead{$\lambda_C$} &
\colhead{$\Delta\lambda_C$} &
\colhead{$f_G$} &
\colhead{$\sigma_G$} &
\colhead{$\sigma_{\rm cont.}$} \\
\colhead{} &
\colhead{(arcsec)} &
\colhead{(arcsec)} &
\colhead{(\AA)} &
\colhead{(\AA)} &
\colhead{($e^{-1}$/hr)} &
\colhead{(\AA)} &
\colhead{($e^{-1}$/hr/res)}
}
\startdata
   1 &    1 & -20.5 & -23.4 & \nodata & \nodata & \nodata & \nodata &     109 \\ 
   2 &    1 & -16.4 & -23.4 & \nodata & \nodata & \nodata & \nodata &      27 \\ 
   3 &    1 & -12.20 & -23.45 & 6637.23 &     0.18  &     728  &    0.59  &      19 \\ 
   4 &    1 & -8.05 & -23.45 & 6637.68 &     0.03  &    1725  &    0.37  &      14 \\ 
   5 &    1 & -3.90 & -23.45 & 6637.67 &     0.03  &    2465  &    0.47  &      14 \\ 
   6 &    1 &  0.25 & -23.45 & 6637.72 &     0.02  &    3854  &    0.49  &      14 \\ 
   7 &    1 &  4.40 & -23.45 & 6637.75 &     0.03  &    2394  &    0.50  &      13 \\ 
   8 &    1 &  35.7 &  47.5 & \nodata & \nodata & \nodata & \nodata &       9 \\ 
   9 &    1 & -22.6 & -19.9 & \nodata & \nodata & \nodata & \nodata &      14 \\ 
  10 &    1 & -18.43 & -19.86 & 6638.13 &     0.11  &     632  &    0.56  &      11 \\ 
  11 &    1 & -14.28 & -19.86 & 6638.07 &     0.09  &     751  &    0.57  &      10 \\ 
  12 &    1 & -10.13 & -19.86 & 6637.59 &     0.04  &    1206  &    0.47  &      10 \\ 
  13 &    1 & -5.98 & -19.86 & 6637.55 &     0.02  &    2756  &    0.46  &       9 \\ 
  14 &    1 & -1.83 & -19.86 & 6637.64 &     0.01  &    5210  &    0.52  &      11 \\ 
  15 &    1 &  2.32 & -19.86 & 6637.73 &     0.01  &    4160  &    0.51  &      10 \\ 
  16 &    1 & -20.5 & -16.3 & \nodata & \nodata & \nodata & \nodata &       9 \\ 
  17 &    1 & -16.35 & -16.26 & 6638.25 &     0.09  &     760  &    0.63  &      10 \\ 
  18 &    1 & -12.20 & -16.26 & 6637.80 &     0.03  &    2256  &    0.49  &      11 \\ 
  19 &    1 & -8.05 & -16.26 & 6637.61 &     0.01  &    5263  &    0.48  &      10 \\ 
  20 &    1 & -3.90 & -16.26 & 6637.51 &     0.01  &    7948  &    0.51  &      11 \\ 
  21 &    1 &  0.25 & -16.26 & 6637.59 &     0.01  &    6561  &    0.54  &      10 \\ 
  22 &    1 &  4.40 & -16.26 & 6637.83 &     0.01  &    3887  &    0.50  &      10 \\ 
  23 &    1 & -22.6 & -12.7 & \nodata & \nodata & \nodata & \nodata &       9 \\ 
  24 &    1 & -18.43 & -12.67 & 6638.57 &     0.17  &     331  &    0.52  &       9 \\ 
  25 &    1 & -14.28 & -12.67 & 6637.82 &     0.02  &    2411  &    0.51  &       8 \\ 
  26 &    1 & -10.13 & -12.67 & 6637.75 &     0.01  &    9654  &    0.50  &      10 \\ 
  27 &    1 & -5.98 & -12.67 & 6637.72 &     0.01  &   16032  &    0.49  &      12 \\ 
  28 &    1 & -1.83 & -12.67 & 6637.64 &     0.01  &   12010  &    0.51  &      12 \\ 
  29 &    1 &  2.32 & -12.67 & 6637.84 &     0.01  &    9437  &    0.50  &      14 \\ 
  30 &    1 & -56.3 &  42.5 & \nodata & \nodata & \nodata & \nodata &       8 \\ 
  31 &    1 & -20.50 & -9.07 & 6638.73 &     0.14  &     527  &    0.64  &       9 \\ 
  32 &    1 & -16.35 & -9.07 & 6638.54 &     0.04  &     960  &    0.50  &       8 \\ 
  33 &    1 & -12.20 & -9.07 & 6637.95 &     0.02  &    4054  &    0.58  &       9 \\ 
  34 &    1 & -3.90 & -9.07 & 6637.86 &     0.01  &   17991  &    0.50  &      15 \\ 
  35 &    1 &  0.25 & -9.07 & 6637.87 &     0.01  &   12811  &    0.51  &      17 \\ 
  36 &    1 &  4.40 & -9.07 & 6638.19 &     0.01  &   12383  &    0.63  &      13 \\ 
  37 &    1 & -18.43 & -5.48 & 6638.84 &     0.03  &    1256  &    0.37  &       8 \\ 
  38 &    1 & -14.28 & -5.48 & 6638.81 &     0.02  &    3654  &    0.50  &       8 \\ 
  39 &    1 & -10.13 & -5.48 & 6638.46 &     0.01  &    8681  &    0.62  &      10 \\ 
  40 &    1 & -5.98 & -5.48 & 6638.20 &     0.01  &   17109  &    0.60  &      16 \\ 
  41 &    1 & -1.83 & -5.48 & 6638.19 &     0.01  &   17713  &    0.64  &      21 \\ 
  42 &    1 & -20.50 & -1.89 & 6638.88 &     0.06  &     598  &    0.41  &       8 \\ 
  43 &    1 & -16.35 & -1.89 & 6638.87 &     0.01  &    4132  &    0.46  &       9 \\ 
  44 &    1 & -12.20 & -1.89 & 6638.89 &     0.01  &    8307  &    0.51  &       9 \\ 
  45 &    1 & -8.05 & -1.89 & 6638.80 &     0.01  &   13259  &    0.67  &      11 \\ 
  46 &    1 & -3.90 & -1.89 & 6638.73 &     0.01  &   17517  &    0.76  &      22 \\
  47 &    1 &  0.25 & -1.89 & 6639.10 &     0.02  &   21608  &    1.04  &      36 \\ 
  48 &    1 &  4.40 & -1.89 & 6639.50 &     0.01  &   19891  &    0.97  &      28 \\ 
  49 &    1 & -22.6 &   1.7 & \nodata & \nodata & \nodata & \nodata &       8 \\ 
  50 &    1 & -18.43 &  1.71 & 6639.04 &     0.04  &     667  &    0.37  &       7 \\ 
  51 &    1 & -14.28 &  1.71 & 6639.17 &     0.01  &    5098  &    0.61  &       8 \\ 
  52 &    1 & -10.13 &  1.71 & 6639.34 &     0.01  &   10843  &    0.76  &      10 \\ 
  53 &    1 & -5.98 &  1.71 & 6639.78 &     0.01  &   17740  &    0.96  &      17 \\ 
  54 &    1 &  2.32 &  1.71 & 6640.30 &     0.01  &   18074  &    0.92  &      32 \\ 
  55 &    1 & -20.5 &   5.3 & \nodata & \nodata & \nodata & \nodata &       8 \\ 
  56 &    1 & -16.35 &  5.30 & 6639.72 &     0.03  &    2233  &    0.59  &       9 \\ 
  57 &    1 & -12.20 &  5.30 & 6639.93 &     0.01  &   11948  &    0.60  &      10 \\ 
  58 &    1 & -8.05 &  5.30 & 6640.27 &     0.01  &   19399  &    0.73  &      12 \\ 
  59 &    1 & -3.90 &  5.30 & 6640.67 &     0.01  &   24179  &    0.68  &      17 \\ 
  60 &    1 &  0.25 &  5.30 & 6640.92 &     0.01  &   17069  &    0.67  &      20 \\ 
  61 &    1 &  4.40 &  5.30 & 6641.05 &     0.01  &    8105  &    0.57  &      16 \\ 
  62 &    1 & -57.3 & -48.5 & \nodata & \nodata & \nodata & \nodata &       6 \\ 
  63 &    1 & -22.6 &   8.9 & \nodata & \nodata & \nodata & \nodata &       9 \\ 
  64 &    1 & -18.4 &   8.9 & \nodata & \nodata & \nodata & \nodata &       9 \\ 
  65 &    1 & -14.28 &  8.90 & 6640.03 &     0.01  &    7204  &    0.55  &       9 \\ 
  66 &    1 & -10.13 &  8.90 & 6640.19 &     0.01  &   14114  &    0.59  &      10 \\ 
  67 &    1 & -5.98 &  8.90 & 6640.78 &     0.01  &   20675  &    0.68  &      12 \\ 
  68 &    1 & -1.83 &  8.90 & 6641.15 &     0.01  &   20157  &    0.63  &      17 \\ 
  69 &    1 &  2.32 &  8.90 & 6641.42 &     0.01  &   10514  &    0.55  &      16 \\ 
  70 &    1 & -20.5 &  12.5 & \nodata & \nodata & \nodata & \nodata &      10 \\ 
  71 &    1 & -16.4 &  12.5 & \nodata & \nodata & \nodata & \nodata &       9 \\ 
  72 &    1 & -12.20 & 12.49 & 6640.40 &     0.02  &    3521  &    0.58  &      10 \\ 
  73 &    1 & -8.05 & 12.49 & 6640.74 &     0.01  &    7907  &    0.61  &      12 \\ 
  74 &    1 & -3.90 & 12.49 & 6641.18 &     0.01  &   11172  &    0.61  &      11 \\ 
  75 &    1 &  0.25 & 12.49 & 6641.55 &     0.01  &    8510  &    0.53  &      14 \\ 
  76 &    1 &  4.40 & 12.49 & 6641.66 &     0.01  &    9271  &    0.51  &      13 \\ 
  77 &    1 & -22.6 &  16.1 & \nodata & \nodata & \nodata & \nodata &       9 \\ 
  78 &    1 & -18.4 &  16.1 & \nodata & \nodata & \nodata & \nodata &      10 \\ 
  79 &    1 & -14.3 &  16.1 & \nodata & \nodata & \nodata & \nodata &      10 \\ 
  80 &    1 & -10.13 & 16.08 & 6640.89 &     0.02  &    2076  &    0.49  &       9 \\ 
  81 &    1 & -5.98 & 16.08 & 6641.10 &     0.01  &    5295  &    0.53  &      11 \\ 
  82 &    1 & -1.83 & 16.08 & 6641.40 &     0.02  &    5337  &    0.57  &      14 \\ 
  83 &    1 &  2.32 & 16.08 & 6641.69 &     0.01  &    7762  &    0.54  &      11 \\ 
  84 &    1 &  33.7 & -54.5 & \nodata & \nodata & \nodata & \nodata &       8 \\ 
  85 &    1 & -20.5 &  19.7 & \nodata & \nodata & \nodata & \nodata &      13 \\ 
  86 &    1 & -16.4 &  19.7 & \nodata & \nodata & \nodata & \nodata &      10 \\ 
  87 &    1 & -8.05 & 19.68 & 6641.19 &     0.02  &    2554  &    0.50  &      11 \\ 
  88 &    1 & -3.90 & 19.68 & 6641.26 &     0.01  &    5194  &    0.49  &      13 \\ 
  89 &    1 &  0.25 & 19.68 & 6641.40 &     0.02  &    3447  &    0.49  &      12 \\ 
  90 &    1 &  4.40 & 19.68 & 6641.64 &     0.03  &    2670  &    0.53  &      12 \\
\enddata
\end{longtable*}

We followed this procedure for the sample of 39 galaxies, or
8280 individual spectra. 6000 (72.5\%) of the spectra 
showed clear emission lines that
were detected and well fit
by the algorithm based on visual inspection. 1875 (22.6\%)
of the spectra had no detectable emission lines. 368 of these 
were sky spectra. Visual inspection flagged 46 (0.6\%) spectra
with potential emission lines that were undetected by the algorithm.
All of these had low S/N and indeed simply may not have met the
S/N criteria. 98 (1.2\%) of the spectra were asymmetric, showing
either a high degree of skew or requiring multiple components. A
single Gaussian profile is not a good fit to these profiles; widths
and centroids are quite possibly in error. They were excluded
from Table 3.  As stated above, these fibers are located very close
to the center. In most cases, they are within a fiber radius of the
center. 48 (0.6\%) of the spectra had fits but were ruled out
by visual inspection. Most of these false positives are weak
cosmic rays or a result of some residual curvature in the spectra
that was not removed through flat-fielding. These are not
included in Table 3. The remaining
213 (2.6\%) of spectra had questionable fits. These algorithm detections
are at a low S/N, and may not have been detected as lines by visual
inspection. Some of these questionable fits may be cosmic rays. Most
are just emission lines which just barely meet the detection threshold.

\subsection{\HI~Reductions}

We reduced our \HI~spectra using the standard Nan\c{c}ay spectral
line reduction packages available at the Nan\c{c}ay site. With this
software we subtracted baselines (generally third order polynomials),
averaged the two receiver polarizations, and applied a
declination-dependent conversion factor to convert from units of
$T_{sys}$ to flux density in mJy.  The $T_{sys}$-to-mJy conversion
factor is determined via a standard calibration relation established
by the Nan\c{c}ay staff through regular monitoring of strong continuum
sources. This procedure yields a calibration accuracy of $\sim$15\%.
In addition, we applied a flux scaling factor of 1.26 to our spectra
based on statistical comparisons (see Matthews \etal 1998, Matthews
\etal 2001) of Nan\c{c}ay survey data of samples of late-type spirals
with past observations of these galaxies made at Nan\c{c}ay and
elsewhere.

From the 23 galaxies observed with Nan\c{c}ay from the two campaigns
(March--August, 1999 and January--June 2001), 18 had sufficient signal
to measure line-widths and integrated \HI~fluxes (Table~4).
The linewidths listed in Table 4 have not been corrected for
redshift stretch. Restframe linewidths should be reduced by
a factor $(1+z)^{-1}$.

\begin{deluxetable*}{cccrrrrrr}
\tabletypesize{\scriptsize}
\tablenum{4}
\tablewidth{0pt}
\tablecaption{\HI\ Parameters from Nan\c{c}ay Observations}
\tablehead{
 \multicolumn{1}{c} {PGC} &  
 \multicolumn{1}{c} {rms\tablenotemark{a}} &
 \multicolumn{1}{c} {S/N\tablenotemark{b}} &
 \multicolumn{1}{c} {\IHI\tablenotemark{c}} &
 \multicolumn{1}{c} {W$_{50}$\tablenotemark{d}} &
 \multicolumn{1}{c} {W$_{20}$\tablenotemark{e}} &
 \multicolumn{1}{c} {$V_{HI}$\tablenotemark{f}} &
 \multicolumn{1}{c} {$\sigma_{V_{HI}}$\tablenotemark{g}} &
 \multicolumn{1}{c} {log \MHI\tablenotemark{h}} \\
 \multicolumn{1}{c} {} &
 \multicolumn{1}{c} {(mJy)} &
 \multicolumn{1}{c} {}  &
 \multicolumn{1}{c} {(Jy km s$^{-1}$)} &
 \multicolumn{1}{c} {(km s$^{-1}$)}  &
 \multicolumn{1}{c} {(km s$^{-1}$)}  &
 \multicolumn{1}{c} {(km s$^{-1}$)} &
 \multicolumn{1}{c} {(km s$^{-1}$)} &
 \multicolumn{1}{c} {(log $M_\odot$)}
}
\startdata
 03512 &  5.82 & \nodata  & \nodata & \nodata  & \nodata  & 5436   & \nodata &\nodata \\
 08941 &  5.31 &  6.69 & 4.72 &  162 &  &  9447 & 5.6 & 10.16\\
 14564 &  2.96 &  26.33 & 10.06 & 176 & 202 & 3483 & 2.2  & 9.71 \\
 15531 &  5.60 &  5.63 & 4.23 & 157  & 197 &  5552 & 12.9 & 9.74 \\
 16274 &  3.74 & \nodata  &\nodata  & \nodata   &\nodata  & 8883 &\nodata &\nodata \\
 19767 &  4.21 & 10.34 & 5.35 &  141 &  164 &  5136 & 5.4 & 9.77 \\
 23598 &  2.80 & 10.41 & 2.57 &  119 &  139 &  7478 & 4.9 & 9.78 \\
 23913 &  3.52 &  34.82 & 4.39 &  33  &  57  &  6330 & 1.6 & 9.87 \\
 24788 &  2.86 & 11.56 & 4.63 &  161 &  194 &  7545 & 5.8 & 10.04 \\
 26140 &  1.79 &   \nodata    &\nodata  &\nodata  & \nodata & 8760 &\nodata  &  \nodata  \\
 28310 &  2.7 & 7.55 & 2.65 & 141 & 153 & 5905 & 5.1 & 9.59 \\
 31159 &  4.9 & 3.26 & 2.32 & 190 & 199 & 5781 & 10.7 & 9.51 \\
 32091 &  3.03 & \nodata & \nodata &\nodata  & \nodata & 2511 &\nodata  & \nodata \\
 32638 &  2.80 &  5.45 & 1.83 &  172 &  229 & 6527 & 15.8 & 9.51 \\
 33465 &  4.41 & \nodata  & \nodata    & \nodata  &\nodata  & 5935  & \nodata &\nodata  \\
 36925 &  7.67 & \nodata  &\nodata & \nodata  &\nodata   & 6826 &\nodata  & \nodata  \\
 38908 &  3.87 &  13.84  & 7.61 &  195 &  221 & 7084 & 4.2 & 10.20 \\
 39728 &  1.97 &  8.81 & 2.08 &  145 &  196 &  2316 & 9.1 & 8.67 \\
 46767 &  2.89 &  8.47 & 4.06 &  276 &  335 & 8065 & 10.4 & 10.07\\
 56010 &  4.80 &  4.78 & 4.80 &   63 &  160 & 4469 & 23.2 & 9.60 \\
 57931 &  6.09 &  & \nodata & \nodata     & \nodata & 9280 &\nodata  &\nodata  \\
 58410 &  5.21 & 5.09 & 4.44 & 256   & 326 &  9055 & 18.9 & 10.18 \\
 72453 &  3.24 & 11.26 & 4.38 &  160 &  194 &  9932 & 6.0 &10.26
\enddata
\tablenotetext{a}{Root mean squared noise level of the spectrum.}
\tablenotetext{b}{Ratio of peak line flux to the rms noise level. During
reduction, data was smoothed to an effective velocity resolution of 9.5 km/s.}
\tablenotetext{c}{Integrated line flux.}
\tablenotetext{d}{Velocity width at 50\% the peak line flux. According
to Fouqu\`e \etal (1990), the uncertainty in $W_{50}$ is
2$\sigma_{V_{HI}}$.}
\tablenotetext{e}{Velocity width at 20\% the peak line flux. According
to Fouqu\`e \etal (1990), the uncertainty in $W_{20}$ is
3$\sigma_{V_{HI}}$.}
\tablenotetext{f}{Heliocentric central radial velocity of a line
profile in the optical convention.}
\tablenotetext{g}{Uncertainty in $V_{HI}$, is $\sigma_{V_{HI}} = 4
R^{1/2}P_W^{1/2} (S/N)^{-1}$ where $R$ is the instrumental resolution
(15.6 km s$^{-1}$), $P_W =(W_{20}-W{50})/2$ (Fouqu\`e \etal 1990).}
\tablenotetext{h}{Total \HI\ mass, $M_{HI} = 2.356\times 10^4 D^2$\IHI
where D is the distance calculated assuming $H_0 = 75$ km s$^{-1}$
Mpc$^{-1}$.}
\end{deluxetable*}

\subsection{\HI~Values from the Literature}

\HI~observations from the literature were drawn from the LEDA
extragalactic database (Paturel 2003) to complement Nan\c{c}ay
observations. Corrections were made to the raw \HI~data to account for
velocity resolution and inhomogeneity of the various original
references to produce velocity widths measured on a common system
using the optical convention ($v=c \Delta\lambda/ \lambda$) at 20\%
and 50\% of the peak flux. These corrected W$_{20}$ and W$_{50}$ line
widths are tabulated for 28 sample galaxies in
Table~5. \HI~observations either from the Nan\c{c}ay observations
described above or from the LEDA database exists for 38 of the 39
sample galaxies.  There are 10 galaxies with W$_{50}$ line widths from
both Nan\c{c}ay observations and the literature, and the measurements
are in good agreement. The mean difference in Nan\c{c}ay versus LEDA
W$_{50}$ line widths is 1.7 km/s; the greatest difference in these
W$_{50}$ values (15.3$\pm12.6$ km/s) is consistent with the errors.
Three of the six galaxies with both Nan\c{c}ay and literature W$_{20}$
line widths show discrepancies greater than 20 km/s.  In these cases,
the relatively low signal to noise ratios of the Nan\c{c}ay
observations for PGC 39728, PGC 46767, and PGC 58410 have led to an
over-estimate of W$_{20}$ (see \S 5.4.1).

\begin{deluxetable}{rrrrr}
\tabletypesize{\scriptsize}
\tablenum{5}
\tablewidth{0pt}
\tablecaption{
\HI~Line-Widths from Paturel \etal (2003)}
\tablehead{
\multicolumn{1}{l}{PGC} &
\multicolumn{1}{l}{W$_{20}$} &
\multicolumn{1}{l}{$\sigma_{{\rm W}_{20}}$} &
\multicolumn{1}{l}{W$_{50}$} &
\multicolumn{1}{l}{$\sigma_{{\rm W}_{50}}$} \\
\multicolumn{1}{l}{} &
\multicolumn{1}{l}{(km s$^{-1}$)} &
\multicolumn{1}{l}{(km s$^{-1}$)} &
\multicolumn{1}{l}{(km s$^{-1}$)} &
\multicolumn{1}{l}{(km s$^{-1}$)} 
}
\startdata
03512  & 204.9      & 10.7       & 182        & 8.1   \\
05345  & 52.49      & 13.32      & 27.05      & 9.7   \\
05673  & 147.75     & 12.85      & 134.05     & 9.7   \\
06855  & 136.5      & 10.7       & 115.7      & 8.1   \\
07826  & 96.08      & 8.32       & 85.4       & 8.24  \\
08941  & 157.22     & 12.85      & 147.32     & 6.86  \\
14564  & 198.93     & 7.39       & 179.64     & 4.62  \\
19767  & 147.9      & 15.86      & 137.14     & 11.94 \\
20938  & \nodata    & \nodata    & 169.61     & 8.5   \\
23333  & 139.12     & 13.28      & 134.42     & 7.1   \\
23598  & \nodata    & \nodata    & 121.37     & 10.49 \\
23913  & \nodata    & \nodata    & 47.37      & 12.51 \\
24788  & \nodata    & \nodata    & 147.76     & 17.61 \\
26517  & 126.5      & 10.7       & 125.7      & 8.1   \\
27792  & \nodata    & \nodata    & 129.87     & 8.47  \\
28401  & 59.5       & 10.7       & 37.7       & 8.1   \\
33465  & 285.9      & 15.86      & 239.14     & 11.94 \\
36925  & 149.9      & 15.86      & 138.14     & 11.94 \\
38268  & 183.33     & 11.97      & 162.92     & 10.02 \\
39728  & 163.28     & 7.08       & 148.96     & 7.46  \\
46767  & 292.9      & 15.86      & 253.14     & 11.94 \\
55750  & 108.05     & 6.41       & 92.61      & 16.26 \\
57931  & 176.41     & 12.54      & 170.19     & 11.78 \\
58410  & 248.9      & 15.86      & 231.32     & 8.68  \\
70962  & 242.35     & 6.03       & 209.09     & 9.73  \\
71106  & 266.78     & 8.46       & 236.86     & 7.34  \\
72144  & \nodata    & \nodata    & 125.31     & 6.97  \\
72453  & 179.2      & 13.28      & 157.3      & 10.02 \\
\enddata
\end{deluxetable}

\section{H$\alpha$ Line-widths}

Creating H$\alpha$ linewidth profiles from DensePak observations is a
computationally straight-forward process, as described below, but it
is necessary to ensure these linewidths are physically meaningful. The
filling factor of DensePak is only 45\% (factoring in the 4 dead
fibers plus the highly vignetted fiber 1). In addition to the low
filling factor, H$\alpha$ may exist at radii beyond the three scale
lengths typically covered by the DensePak observations. Further, since
the sample is nearly face-on, the ``turbulent'' component of the
spatially-integrated line-width is critical to understand.  Therefore,
we first characterize the spatially-resolved H$\alpha$ line-widths, as
observed with DensePak. After describing the construction of the
DensePak H$\alpha$ linewidth profiles, we compare the DensePak
H$\alpha$ linewidth profile to one generated from narrow-band
H$\alpha$ imaging. We also comment briefly on the shape and width of
the \HI~and H$\alpha$ profiles, with a more detailed analysis
presented in the relevant forthcoming paper in this series.

\subsection{Spatially-Resolved DensePak Line-widths}

The projected H$\alpha$ velocity widths from individual fibers vary
little with radius, when considering the sample taken as a
whole. Variations do exist, however, between and within galaxies;
these variations are explored in a future paper. The sample-mean
near-constancy with galactic radius is similar to what has been found
for individual, nearly face-on disk galaxies in neutral gas (NGC 628,
NGC 1058 and NGC 3938; van der Kruit \& Shostak 1982, 1984; Dickey
\etal 1990), molecular gas (NGC 628 and NGC 3938; Combes \& Becquaert
1997), and ionized gas (NGC 3938 and NGC 5668; Jim\'{e}nez-Vicente
\etal 1999; Jim\'{e}nez-Vicente \& Battaner 2000). The reasons for
this gross similarity of the near-constancy with radius of 
different phases of the ISM 
is outside the scope of this paper, but we do
note that some radial dependence in the neutral gas is observed in
some galaxies (Shostak \& van der Kruit 1984; Kamphuis \& Sancisi
1993), and variations are also seen in the ionized gas in some cases
in our sample, albeit over different radial ranges. Our distribution
median, out of 6000 fiber samples with emission lines well
characterized by a single Gaussian across 39 galaxies in our sample is
$17.9\pm^{3.5}_{2.9}$$\pm^{8.1}_{5.7}$ km s$^{-1}$, where the two sets
of ``errors'' include 50\% and 80\% of the distribution. These values
are corrected for instrumental broadening only. They are consistent in
value and distribution for what has been observed in two other, nearby
spirals NGC 3938 and NGC 5668 (Jim\'{e}nez-Vicente \etal 1999,
2000), and are roughly two to three times the dispersion measured in 
either neutral (van der Kruit \& Shostak 1982, 1984) or molecular gas
(Combes \& Becquaert 1987) for relatively small samples of galaxies.
The tightness of our sample distribution is shown in Figure
6. The largest values which skew the distribution slightly are seen
primarily at the center of galaxies with steep inner rotation
curves. While beam smearing appears to affect these values, the effect
is small, correctable, and does not adversely influence our results
here.

\begin{figure}
\vbox to 3.3in{\rule{0pt}{3.3in}}
\includegraphics{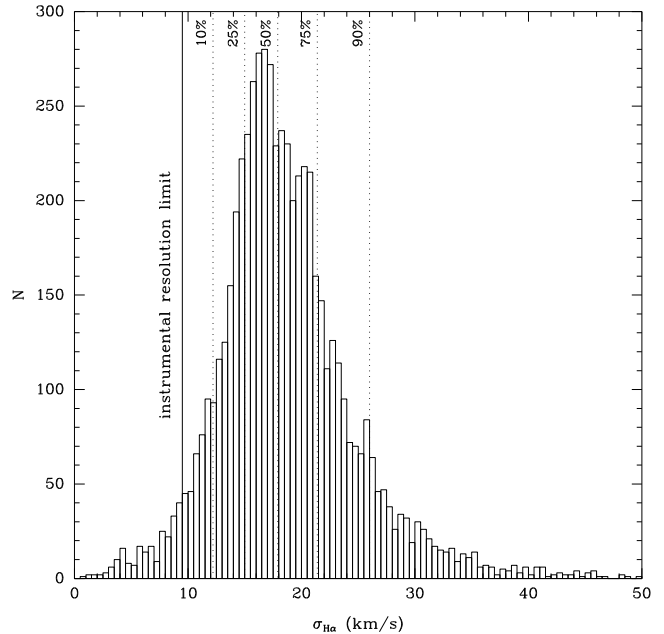}
\caption{Distribution of H$\alpha$ emission-line velocity widths
($\sigma$), as observed for the entire sample of 39 galaxies,
corrected for instrumental broadening. The instrumental resolution
limit and integral distribution percentiles are indicated. The
instrumental resolution is well below the peak of the observed
distribution.}
\end{figure}

\subsection{Spatially-Integrated DensePak Line-widths}

After taking the steps as described in \S 4.1 of extracting the
spectra and fitting Gaussian line-profiles, we generated ``spatially
integrated'' line-widths -- what we refer to here also as ``synthetic
line profiles.'' We coadded either the individual extracted spectra
(channel by channel), or we coadded the individual fitted Gaussian
line-profiles for those spectra where fits were possible.  When
combining these data or fitted profiles, we maintained a count of the
number of independent fiber positions. An on-sky location was assigned
to each fiber based on the DensePak geometry and telescope offsets
determined via velocity-field fitting (presented in a future paper).
The location of fibers from different pointings were checked for
overlap. If there was an overlap between fibers, the relevant fiber
flux was scaled such that the combined spectra gave equivalent spatial
sampling to non-overlapping areas. Specifically, for each fiber we
determined the number of fibers in a hexagonal area inscribed within a
6.3 arcsec radius circle centered on the fiber, and compared this to
the expected number in non-overlapping regions (e.g., 7 fibers for the
middle of the array).  We did not include spectra from fiber 1, as
they suffered from extremely low signal to noise.

\subsection{Spatially-Integrated Line-widths from Narrow-band Imaging}

To test whether the H$\alpha$ linewidth profiles derived from DensePak
observations were valid representations of the matter and velocity
distribution of the ionized gas, we constructed a linewidth profile
from the narrow-band H$\alpha$ image of one galaxy in our sample, PGC
38268, for comparison.

\subsubsection{Observations}

PGC 38268 was observed in narrow-band filters using the SPIcam camera
at the Apache Point Observatory (APO) 3.5m telescope on April 11,
2002. SPIcam is an imager equipped with a $2048 \times 2048$ pixel
SITe back-illuminated CCD that has a $4.8\arcmin$ field of view. In
the standard mode of operation ($2 \times 2$-binning) SPIcam has an
effective plate scale of $0.281\arcsec$/pixel. Rest-frame H$\alpha$
and a 6629\AA~off-band filter come from a set owned by
P. Hodge. SPIcam is designed to accommodate $3 \times 3$-inch filters,
so the $2 \times 2$-inch narrow-band filters consequently reduce
SPIcam's effective field of view (FOV).

The recession velocity of PGC 38268 shifted the H$\alpha$ emission
into the 6629\AA~narrow-band filter. The H$\alpha$ and 6629\AA~filters
make an ideal pair of filters for narrow-band imaging because the
filters' transmission curves are virtually the same in width and
shape.  Since the night was not photometric, no attempt was made to
collect flux standards. Exposures in each band totaled 10 minutes.

\subsubsection{Reductions}

We used standard IRAF tasks to process images (overscan correction,
bias subtraction, and flat-fielding). Images were trimmed to include
only the un-vignetted FOV.  The standard IRAF cosmic ray cleaning task
was augmented using information from neighboring pixels, thus
enhancing the cleaning of extended cosmic ray events and detector flaws.

Once individual data frames were fully processed, the sky background
was calculated using an iteratively clipped mean and subtracted from
all frames. On and off-band images were registered using 5 field
stars. Since the ratio of fluxes in on versus off-band for these stars
were in agreement with the predicted ratio based on the filter
transmission curves, we used the ratio expected from the transmission
curves to scale the images and subtract frames.  The H$\alpha$ narrow
band image of PGC 38268 with the overlay of DensePak fibers in which
H$\alpha$ was detected is shown in the first panel of Figure 7. While
DensePak does not detect all the H$\alpha$ flux because the the two
clumps of H$\alpha$ to the SW and the faint clumps to the N were not
covered by the DensePak array, DensePak is able to detect very faint
levels of H$\alpha$ emission which are not visible in the image.

\begin{figure*}
\vbox to 3.1in{\rule{0pt}{3.1in}}
\includegraphics{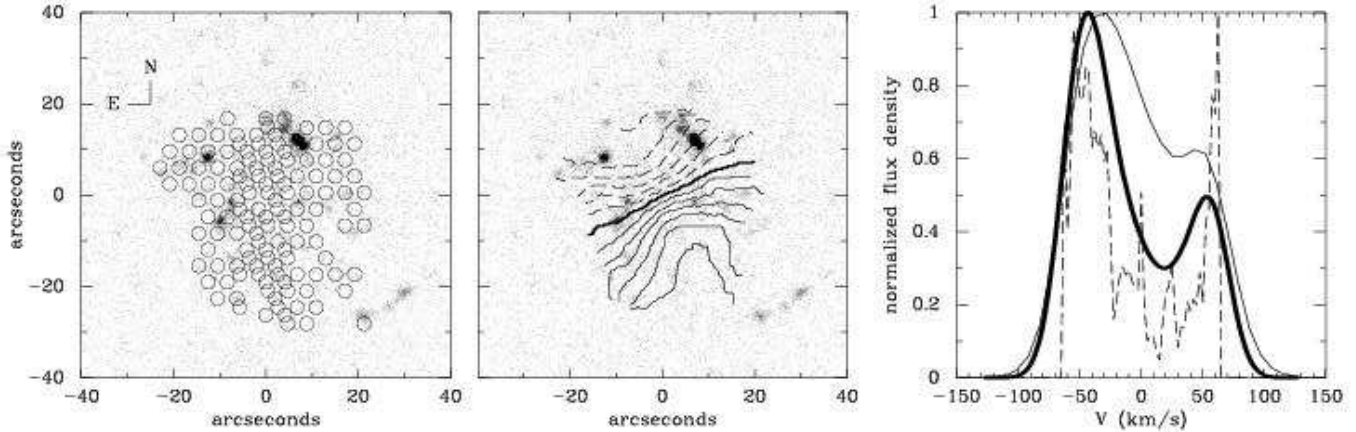}
\caption{{\bf Left Panel:} Narrowband H$\alpha$ image of PGC 38268
with overlay of the DensePak fiber locations in which H$\alpha$ was
detected. Aside from the H$\alpha$ flux missed at large radii due to
the limited field of view of Densepak, the IFU successfully detected
H$\alpha$ across the galaxy even at locations in which the imaging
data does not yield discrete H$\alpha$ detections.  {\bf Center Panel:} The
DensePak H$\alpha$ velocity field overlayed on the narrowband
H$\alpha$ image of PGC 38268. {\bf Right Panel:} Normalized flux
density of the PGC 38268 
H$\alpha$ line profile. Jagged dashed line is
the raw line profile obtained by convolving the velocity field model
and the H$\alpha$ flux from the narrowband image.
The heavy solid line is this same profile convolved with a Gaussian
with 20 km s$^{-1}$ width (which corresponds to the measured mean emission 
line width before instrumental correction was applied). 
The thin solid line is the DensePak line profile obtained by summing the
individual Gaussian fits. The simulated profile has virtually the
same width as the data and shows a qualitatively similar asymmetry
profile.}\end{figure*}

\subsubsection{A Comparison of Line Profiles}

We use this H$\alpha$ narrow-band image of PGC 38268 to produce a
linewidth profile. To do this, we assign a velocity to each position
in the image.  Following the fitting procedure outlined in Andersen
\etal (2001), a best fit velocity field was determined from the
DensePak data. The second panel of Figure 7 shows the observed
velocity field from the DensePak data overlaid on the narrow-band
image of PGC 38268. The best fit analytic model to this observed
velocity field was then used to assign a velocity to each pixel of the
narrow-band image. The resultant profile (dashed line of Figure 7) was
then smoothed by a 20 km s$^{-1}$ width Gaussian profile (solid line
of Figure 7).  This smoothing corresponds to the median measured
H$\alpha$ line-width in spectra of individual DensePak fibers.

When the linewidth derived from the DensePak data if overlaid on this
image (dotted line of Figure 7), good agreement is observed between
the smoothed profile and the DensePak profile. W$_{20}$, the velocity
width measured at 20\% of the peak velocity, when measured from the
profile derived from the H$\alpha$ image is 152 km s$^{-1}$ and when
measured from the profile created directly from the DensePak spectra
is 162 km s$^{-1}$. A difference of 5 km s$^{-1}$ in the width of the
smoothing Gaussian used to smooth the line profile can account for
this difference. The more striking discrepancy is between the peak
flux levels in the two halves of the profile.  The ratio of fluxes
from the DensePak data is 0.6 while the ratio is 0.5 when measured
from the H$\alpha$ image. This agrees qualitatively with the relative
positions of H$\alpha$ emission and the DensePak footprint: More
H$\alpha$ flux falls between fibers in the receding (Northern) half of
the velocity field.  Finally, the difference in velocity between the
two peaks is slightly greater in the data from the H$\alpha$
imaging. This is consistent with the non-detection of H$\alpha$ at
large radii and velocities in the DensePak data.

Overall, however, we conclude the DensePak-derived integrated
line-width is likely an excellent approximation.

\subsection{Synthetic H$\alpha$ Line-Profiles}

We present these synthetic line-profiles for both the raw data and the
summed, best-fit Gaussian profiles in Figure 8, along with an overlay
of the DensePak observations on the $R$-band images.  As can be seen
by direct visual inspection, the line profile width and shape is very
similar for the raw and Gaussian summations.  The smoothing of the
data with the Gaussion summation does not quantitatively alter
low-order moments of the integrated line-profiles. For example, we
find the variation of the skew, or lopsidedness, of the line profile
between raw and Gaussion summations is less than the observational
error, and are identical in the mean.

A comparison between the \HI~ and H$\alpha$ profiles are also
presented in Figure 8, where the data are available. In general there
is very good agreement between the widths of the \HI~and H$\alpha$
profiles. Using the Gaussian fits to the individual H$\alpha$ spectra,
we constructed a width by finding the maximum difference in centroid
velocities. We compared this width to the 31 \HI~W$_{20}$ measures and
found a constant offset between the relations. The \HI~line widths are
23$\pm20$ km/s broader than the H$\alpha$ velocity ``widths'' as
defined above, roughly comparable to the $\sim$18 km/s
spatially-resolved line-widths of the H$\alpha$ emission lines.  Only
two of these galaxies, PGC 56010 and PGC 70962, showed significantly
larger differences in velocity widths. As noted below, only half the
disk of PGC 56010 is covered with DensePak pointings.  Similarly, we
only cover the inner starbursting region of PGC 70962, but we suspect
the width of the \HI~profile may be dominated by the especially broad
emission lines observed in the H$\alpha$ spectra and may not reflect
the kinematics of the galaxy.  Further comparisons of \HI~ and
H$\alpha$ line-widths are in a forthcoming paper.

\subsubsection{Notes to Individual Galaxies}

For galaxies with both DensePak H$\alpha$ and Nan\c{c}ay \HI~line-profiles
shown in Figure 8, we normalized the peak fluxes of the
H$\alpha$ line-profiles generated from the sum of individual Gaussian
fits to the \HI~line profile peak flux for comparison.  Here, we take
note of the features observed in the line-widths for galaxies with
both DensePak H$\alpha$ and Nan\c{c}ay \HI~data:

\begin{description}

\item[{\bf PGC 3512:}] No \HI~flux from the source was detected.

\item[{\bf PGC 08941:}] Nan\c{c}ay observations were severely
affected by radio interference, but no \HI~flux was detected at the
recession velocity of the source.

\item[{\bf PGC 14564:}] \HI~and H$\alpha$ line-profiles are very
similar. The only notable difference is the slightly broader width of
the H$\alpha$ profile.

\item[{\bf PGC 15531:}] S/N of the \HI~linewidth is low, but the
shape of the line-profiles are very similar.

\item[{\bf PGC 16274:}] Nan\c{c}ay observations had a poor
base-line, making a comparison between H$\alpha$ and \HI~line-widths
difficult.

\item[{\bf PGC 19767:}] H$\alpha$ data exhibits a strong asymmetry
in the line profile that is not observed in the \HI~data.

\item[{\bf PGC 23598:}] H$\alpha$ and \HI~line-widths are equivalent
and both profiles are asymmetric. The detailed shapes of the profiles,
however, are markedly different.

\item[{\bf PGC 23913:}] Neither line profile shows evidence of
significant rotation.

\item[{\bf PGC 24788:}] \HI~data has a slightly broader line width and
is double-peaked while the H$\alpha$ profile is not.

\item[{\bf PGC 26140:}] There is a strong mismatch between H$\alpha$
and \HI~line profiles, with little \HI~signal at the source
redshift, but a possible, narrow peak blue-shifted by 300-350 km
s$^{-1}$.  Within the Nan\c{c}ay beam lies PGC 26140 (\am{2}{4}
separation), an Elliptical with redshift (V=8570$\pm$33 km s$^{-1}$;
LEDA), size, and magnitude similar to those the face-on spiral
(V=8760$\pm$141 km s$^{-1}$). Though our central \HI~velocity (8575
km s$^{-1}$) is closer to the optical velocity of the Elliptical, we
expected it to be gas-poor and our 21cm line detection should not be
confused by it. PGC 26140 was not detected at Arecibo (Giovanardi \&
Salpeter 1985) with a quoted 3$\sigma$ upper limit of 1.6 Jy km
s$^{-1}$ for a 300 km s$^{-1}$ wide flat-topped profile, considerable
lower than our detected integrated line flux of 3.1 Jy km
s$^{-1}$. We can conclude that the \HI~mass for this galaxy
lies far below the predicted mass, $\sim10.3$ log \MHI , based
on its luminosity and type (Roberts \& Haynes 1994).

\item[{\bf PGC 28310:}] H$\alpha$ and \HI~line-widths are
equivalent and both profiles are asymmetric. The detailed shapes of
the profiles, however, are markedly different.

\item[{\bf PGC 31159:}] \HI~and H$\alpha$ line-profiles are equivalent,
but the \HI~profile is double-peaked while the H$\alpha$ is not.

\item[{\bf PGC 32091:}] No \HI~flux was detected.

\item[{\bf PGC 32638:}] Nan\c{c}ay observations have a poor
base-line, making a comparison between H$\alpha$ and \HI~line-widths
difficult.

\item[{\bf PGC 33465:}] Nan\c{c}ay observations have a poor
base-line, making a comparison between H$\alpha$ and \HI~line-widths
difficult.

\item[{\bf PGC 36925:}] No \HI~flux from the source was detected.

\item[{\bf PGC 38908:}] \HI~and H$\alpha$ line-widths are very similar.
The only notable difference is the slightly broader width of the H$\alpha$
profile.

\item[{\bf PGC 39728:}] \HI~and H$\alpha$ line-widths are very similar.
The \HI~profile displays a strong asymmetry that is not observed in the
H$\alpha$ profile.

\item[{\bf PGC 46767:}] \HI~profile has a low S/N, but the H$\alpha$
and \HI~profiles and profile-asymmetries are similar
nonetheless. W$_{20}$ measured from Nan\c{c}ay profile is
significantly larger than one would expect from the profile due to the
baseline noise.

\item[{\bf PGC 55750:}] The spatial coverage of the DensePak pointings
were insufficient to cover the disk of the galaxy (Observations were
performed for only 2 of 3 planned pointings).

\item[{\bf PGC 56010:}] Neither line-profiles show much evidence for
rotation. The spatial coverage of the DensePak pointings were
insufficient to cover the disk of the galaxy (Observations were
performed for only 2 of 3 planned pointings).

\item[{\bf PGC 57931:}] Nan\c{c}ay observations have a poor
base-line, making a comparison between H$\alpha$ and \HI~line-widths
difficult.

\item[{\bf PGC 58410:}] \HI~and H$\alpha$ line-profiles are very
similar.

\item[{\bf PGC 72453:}] H$\alpha$ and \HI~line-widths are equivalent
and both profiles are asymmetric. The detailed shapes of the profiles,
however, are markedly different.

\end{description}

\section{Summary and Discussion}

We used the Principal Galaxy Catalog to choose a sample of 39 disk
galaxies which span a range in luminosity, surface brightness and
type. For this sample, we gathered H$\alpha$ emission-line data for 39
nearly face--on spiral galaxies with the DensePak IFU on WIYN. These
observations provide a fairly uniform spatial sampling of the ionized
gas out to radii of approximately 3 scale lengths with a spectral
resolution sufficient to yield emission-line velocities with a
centroiding accuracy of $\sim 2$ km s$^{-1}$.

For one of these galaxies, PGC 38268, we collected a deep narrow-band
H$\alpha$ image. By comparing the DensePak data to the narrow-band
image, we found that the 2.8$^{\prime\prime}$ DensePak fibers and
3$\times$ longer exposures enabled us to detect discrete line-emission
over a much larger filling factor than apparent in the narrow-band
images. Despite the different detection thresholds, we used the PGC
38268 narrow-band image convolved with a model of the velocity field
obtained from the DensePak data to construct a spatially-integrated,
H$\alpha$ line-profile which accounts for all H$\alpha$ gas in the
galaxy. A line profile constructed using only the DensePak data,
despite not accounting for all the ionized gas, still was able to
produce a comparable, integrated line-width. Qualitatively, the two
H$\alpha$ emission-line-profiles exhibited the same width {\it and}
asymmetry. Having demonstrated that DensePak observations can be used to
produce a realistic line profile for PGC 38268, we generated spatially
integrated H$\alpha$ line-widths for all galaxies in our sample.

We have also measured the spatially-resolved velocity dispersion for
ionized gas in our sample. For the 6000 IFU spectra which were
well-fit by a single Gaussian (and therefore less sensitive to beam
smearing), we remark that under the assumption that the thermal
line-width for ionized gas at 10$^4$ $K$ is $\sigma = 9.1$ km s$^{-1}$
(Osterbrock 1989), and the intrinsic line-width is $\sigma = 3$ km
s$^{-1}$, then following the quadrature formalism of
Jim\'{e}nez-Vicente \etal (1999), the turbulent motions for \HII\ in our
sample ranges from 7.7 to 24.2 km s$^{-1}$ (80\%). This range may seem
large, but is in keeping with the amplitude and range seen in
Fabry-Perot studies of much smaller samples of galaxies
(cf. Jim\'{e}nez-Vicente et al. 1999). Perhaps surprisingly,
significant variations and similar amplitudes are seen in the
molecular and neutral gas -- albeit in small samples. For instance,
Combes \& Becquaert (1997) find 6 and 8.5 km s$^{-1}$ dispersions for
CO in NGC 628 and NGC 3938, respectively. While van der Kruit \&
Shostak (1982, 1984) have remarked on the constancy of \HI\ velocity
dispersions in disks ranging from 7 to 10 km s$^{-1}$, this is based
on radio-synthesis observations of three galaxies. One of these
galaxies, NGC 1058, was re-observed by Dickey \etal (1990) at higher
spectral resolution. They found a lower \HI\ velocity dispersion of 5.7
km s$^{-1}$.  Both Shostak \& van der Kruit (1984) and Kamphuis \&
Sancisi (1993) have also noted \HI\ velocity dispersions in two
late-type, nearly face-on disk decrease from 10-13 km/s in the inner
disk (what Kamphuis \& Sancisi call the ``turbulent regions of the
optical disks'') to 6-8 km/s in the outer disk.  While the physical
mechanisms driving the velocity dispersions of the different ISM
phases may indeed be different (cf. McKee \& Ostriker 1977, and Jog \&
Ostriker 1988), our primary points here is the variation of turbulent
motions from galaxy to galaxy and within galaxies in all phases may be
more significant than previously supposed.  We will return to this
issue in the context of the ionized gas in latter papers in this
series.

We gathered available, spatially-integrated \HI~line-profiles from
the literature and augmented them with new 21-cm observations from the
Nan\c{c}ay radio telescope. We have presented new \HI~line-profiles
for 23 galaxies, along with measures of their line-widths and
\HI~fluxes. These data allow us to compare line-profiles of ionized
and neutral hydrogen.  For galaxies with sufficient S/N in their
\HI~profiles, we found excellent agreement between the width of the
profiles indicating that we are sampling the flat portion of the
velocity fields in H$\alpha$ for most of this sample. However, the
sample of \HI~and H$\alpha$ line-profiles showed significantly
different profile shapes indicating that we are observing different
distributions in the ionized versus neutral gas.

We have shown that integral field spectroscopy does an admirable job 
of obtaining integrated line-widths. Line-profiles from Nan\c{c}ay 
have S/N ratios far less than the S/N ratios for line-profiles 
constructed from DensePak data. Yet the Nan\c{c}ay radio telescope has 
a collecting area more than 800 times greater than WIYN, and we used 
an average on-source integration two times longer per galaxy in \HI\ 
than for H$\alpha$. Even if we ignore the different signal-to-noise 
ratios of the lines, and account for the fact that radio telescopes 
are roughly 100 times less expensive per m$^2$ than optical telescopes 
(of comparable steerability; van Belle \etal 2004), and that \HI\ 
observations can be carried out during daylight hours, the enhanced 
cost-effectiveness of DensePak-to-Nan\c{c}ay observations is greater 
than a factor of 5. Since our data was taken, Nan\c{c}ay has undergone 
an extensive upgrade which has increased the sensitivity by a factor 
of $\sim 5$, which brings the cost effectiveness roughly to 
unity. High signal-to-noise ratio \HI\ line profile observations using 
the Green Bank 43m and Arecibo 305m radio telescopes have comparable 
cost effectiveness to our DensePak observations (e.g. Haynes \etal 
1998; Haynes \etal 1999). Yet the true forte of H$\alpha$ integral 
field spectroscopy is the high spatial resolution of resolved velocity 
fields and the simultaneous observations of multiple atomic species. 
Typical radio-synthesis \HI\ maps require 10 to 20 hours of 
integration on Westerbork or the VLA, with most radio observations 
using beam sizes of 15 arcsec or greater. This is characteristically 
five times longer integrations at 5 times lower resolution than 
DensePak. \HI\ and H$\alpha$ observations probe different ions and 
physical scales, and so remain complementary. However, with the 
relatively recent advent of bi-dimensional optical spectroscopy with 
DensePak and related IFUs, it appears that efficient kinematic studies 
within the inner (optically bright) regions of galaxies are now 
squarely in the optical regime. 

\acknowledgements

We thank D. Zucker and H. Lee for providing the narrow-band H$\alpha$
imaging data for PGC 38268. This research was supported by NSF grants
AST-9970780 and AST-0307417. DRA also acknowledges the support of the
NASA Graduate Student Research Program. MAB gratefully acknowledges
the hospitality he enjoyed during his stay at the University of
Toronto, during which time the work was completed.


\bibliographystyle{apj}

\clearpage

\begin{figure*}
\vbox to 8.6in{\rule{0pt}{8.6in}}
\includegraphics{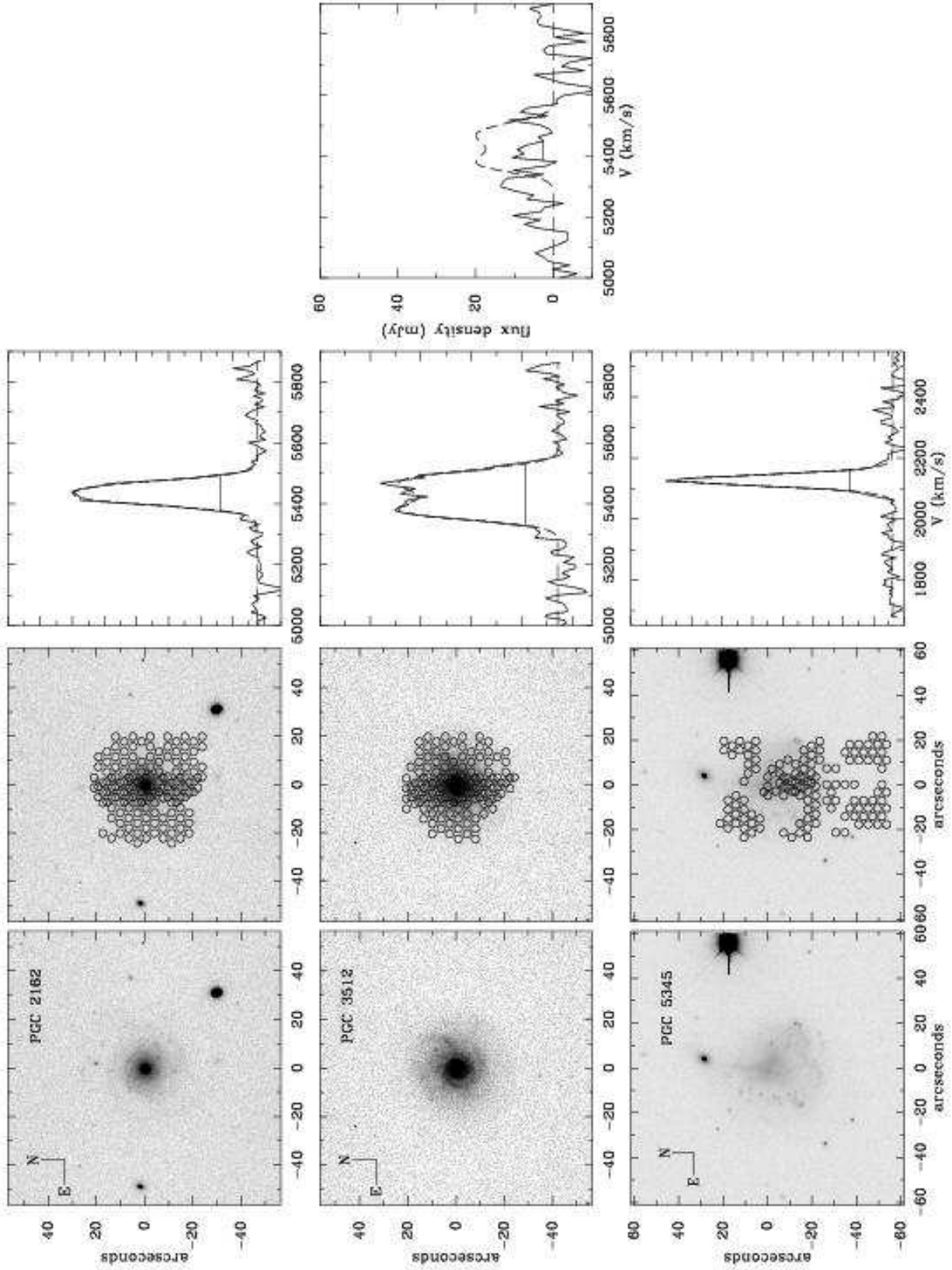}
\caption{{\bf Leftmost Column:} Images of the 39 sample galaxies with
overlays of the DensePak fibers in which H$\alpha$ is detected. {\bf
Central Column:} H$\alpha$ velocity profiles from the DensePak
observations. The H$\alpha$ flux densities are plotted against
heliocentric velocities. No units are given on the flux densities
because nights were generally clear but not photometric, nor were
flux-standards observed. The smoother profiles are sums of Gaussian
fits to individual fiber H$\alpha$ profiles. These ``Sum of Gaussian''
profiles agree well with the raw data. Flux from fibers with centers
that fall within a fiber radii are averaged. W$_{20}$, the width of the
H$\alpha$ lines at 20\% the peak flux is illustrated by a horizontal
line in the plots. {\bf Rightmost Column:} The \HI~flux densities
from Nan\c{c}ay 21cm observations (where available) versus
heliocentric velocities. The smooth curve is again the H$\alpha$ ``Sum
of Gaussian'' profile scaled to the \HI~peak flux.  W$_{20}$, the width
of the \HI~lines at 20\% the peak flux is illustrated by a horizontal
line in the plots.}
\end{figure*}

\clearpage

\begin{figure*}
\vbox to 7.5in{\rule{0pt}{7.5in}}
\includegraphics{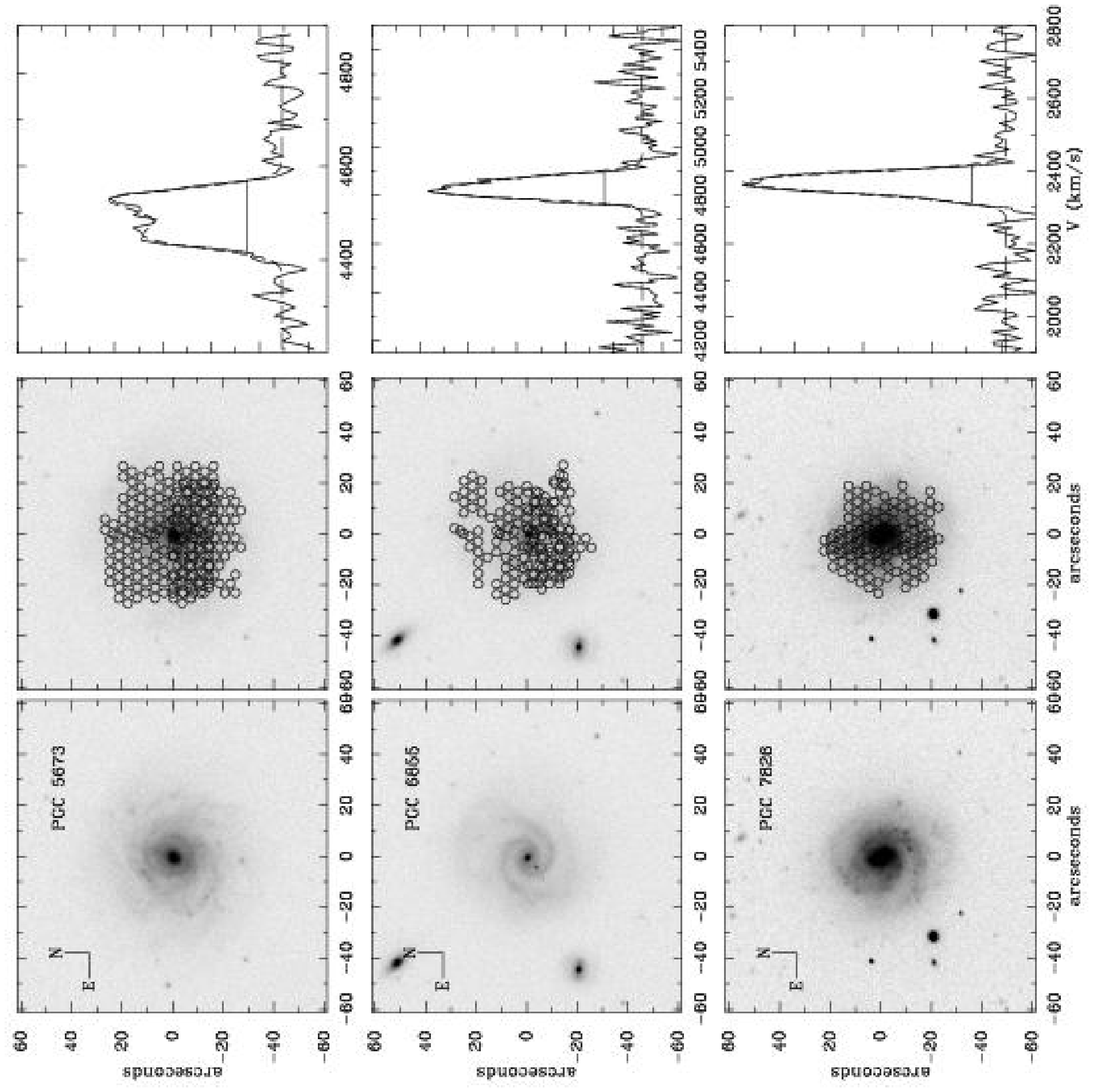}
\end{figure*}

\clearpage

\begin{figure*}
\vbox to 7.5in{\rule{0pt}{7.5in}}
\includegraphics{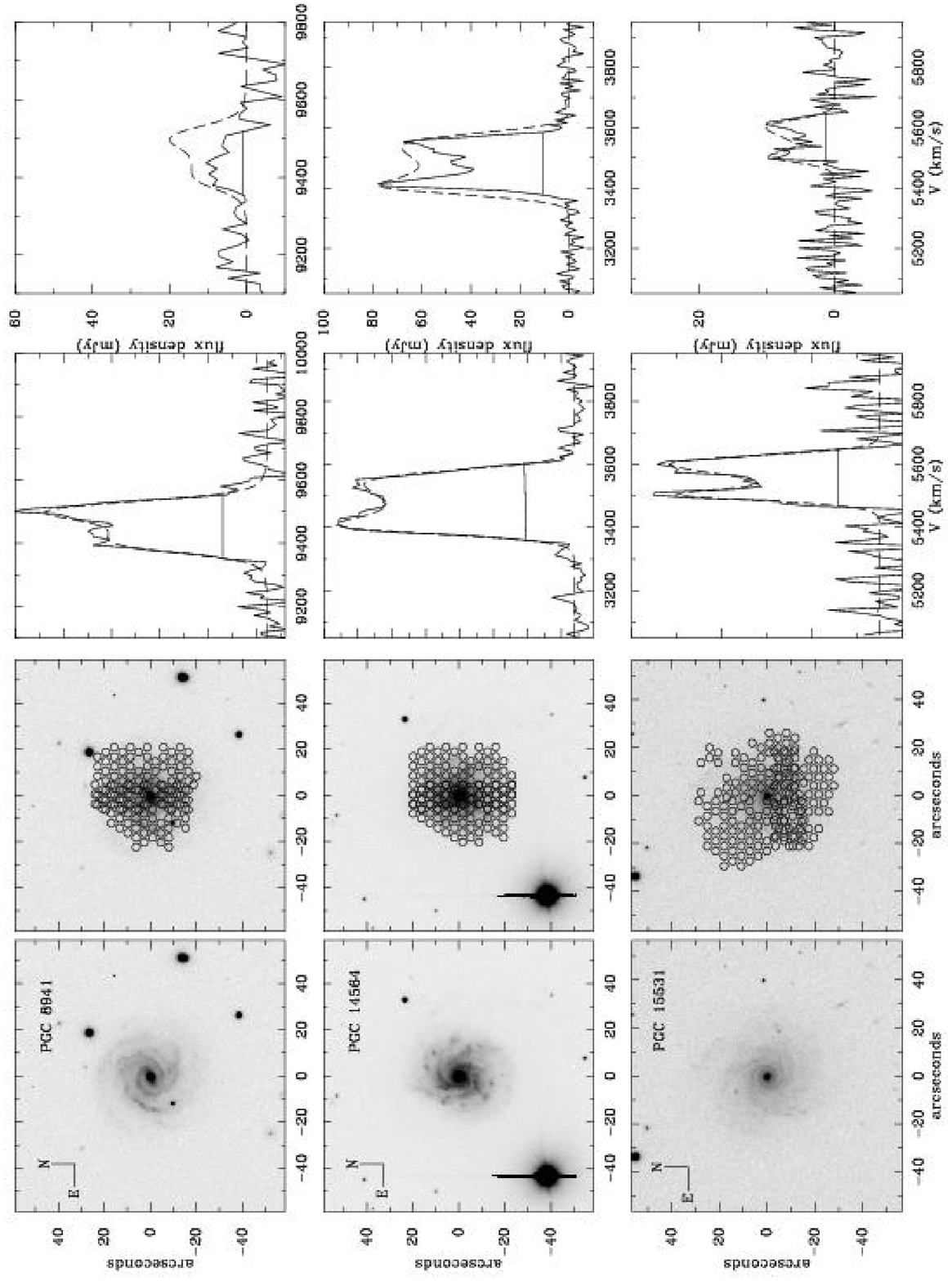}
\end{figure*}

\clearpage

\begin{figure*}
\vbox to 7.5in{\rule{0pt}{7.5in}}
\includegraphics{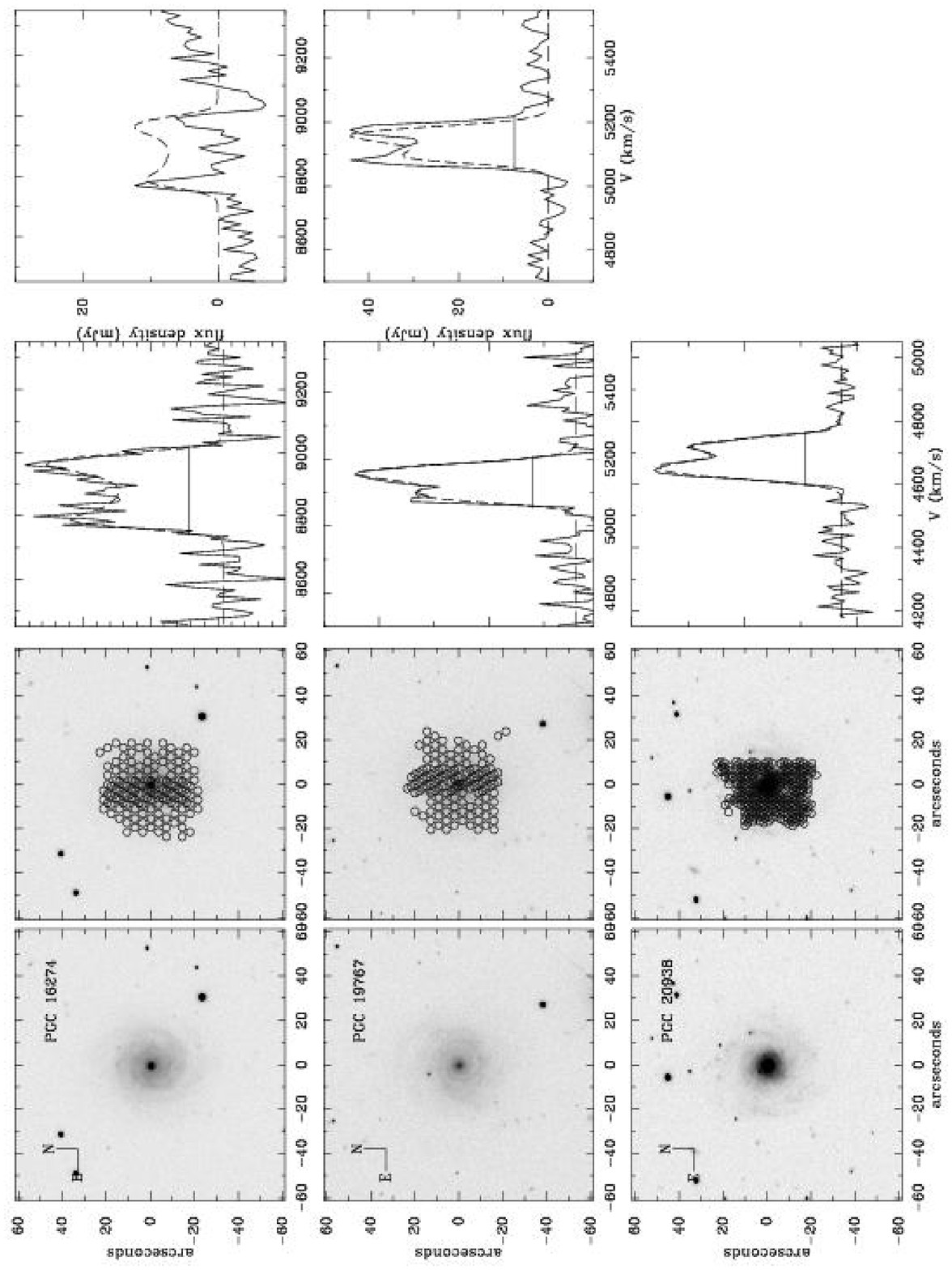}
\end{figure*}

\clearpage

\begin{figure*}
\vbox to 7.5in{\rule{0pt}{7.5in}}
\includegraphics{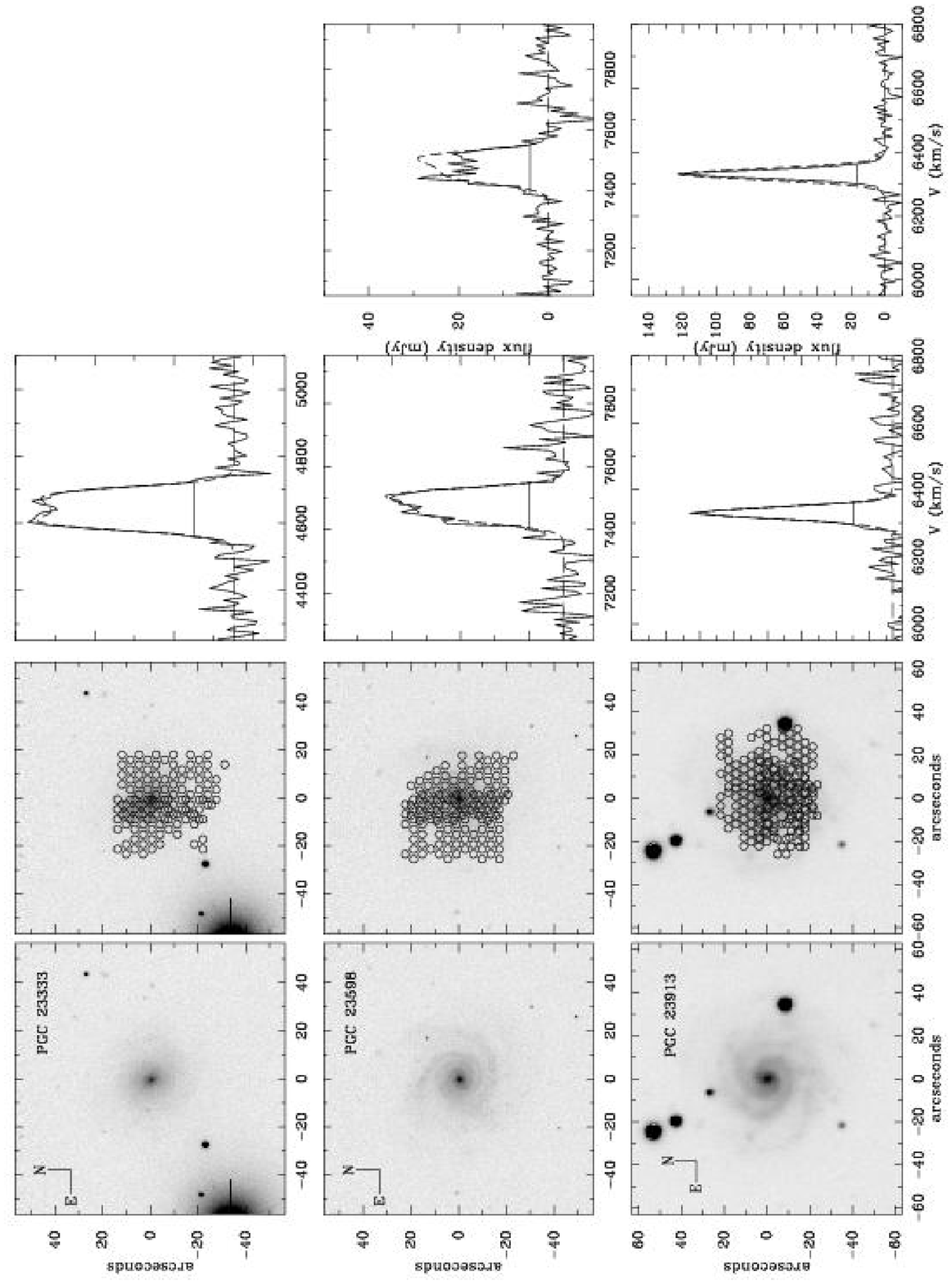}
\end{figure*}

\clearpage

\begin{figure*}
\vbox to 7.5in{\rule{0pt}{7.5in}}
\includegraphics{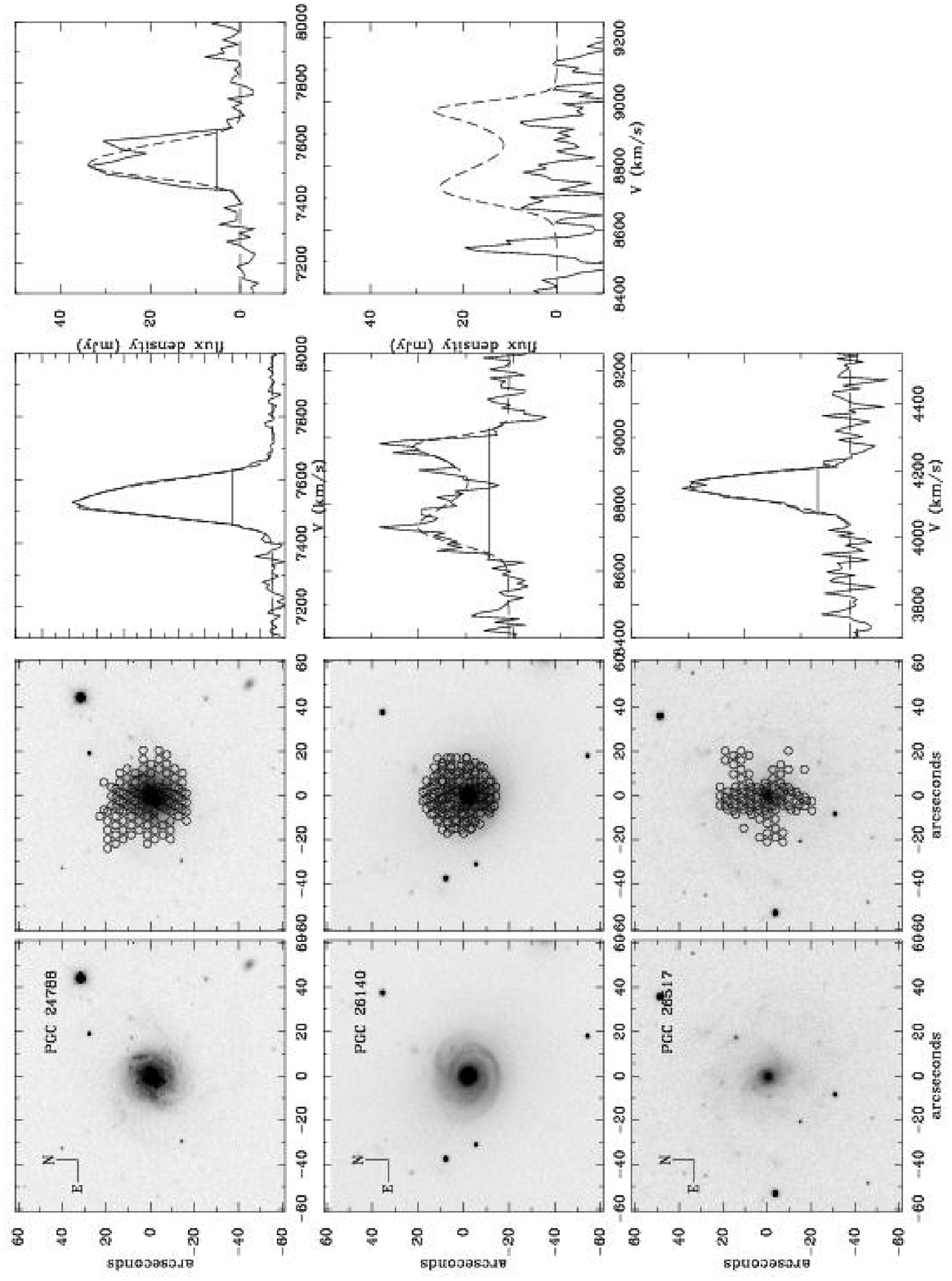}
\end{figure*}

\clearpage

\begin{figure*}
\vbox to 7.5in{\rule{0pt}{7.5in}}
\includegraphics{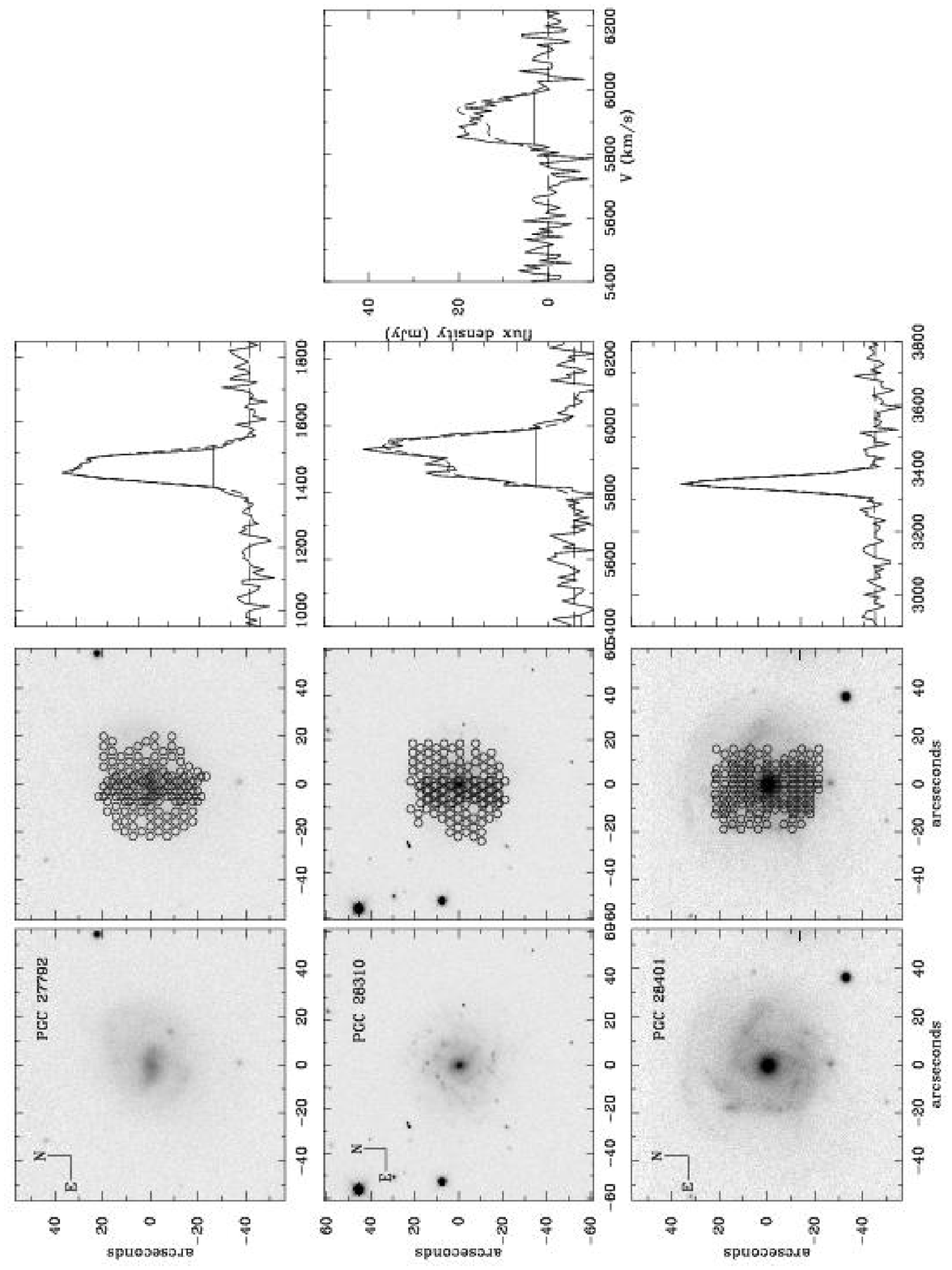}
\end{figure*}

\clearpage

\begin{figure*}
\vbox to 7.5in{\rule{0pt}{7.5in}}
\includegraphics{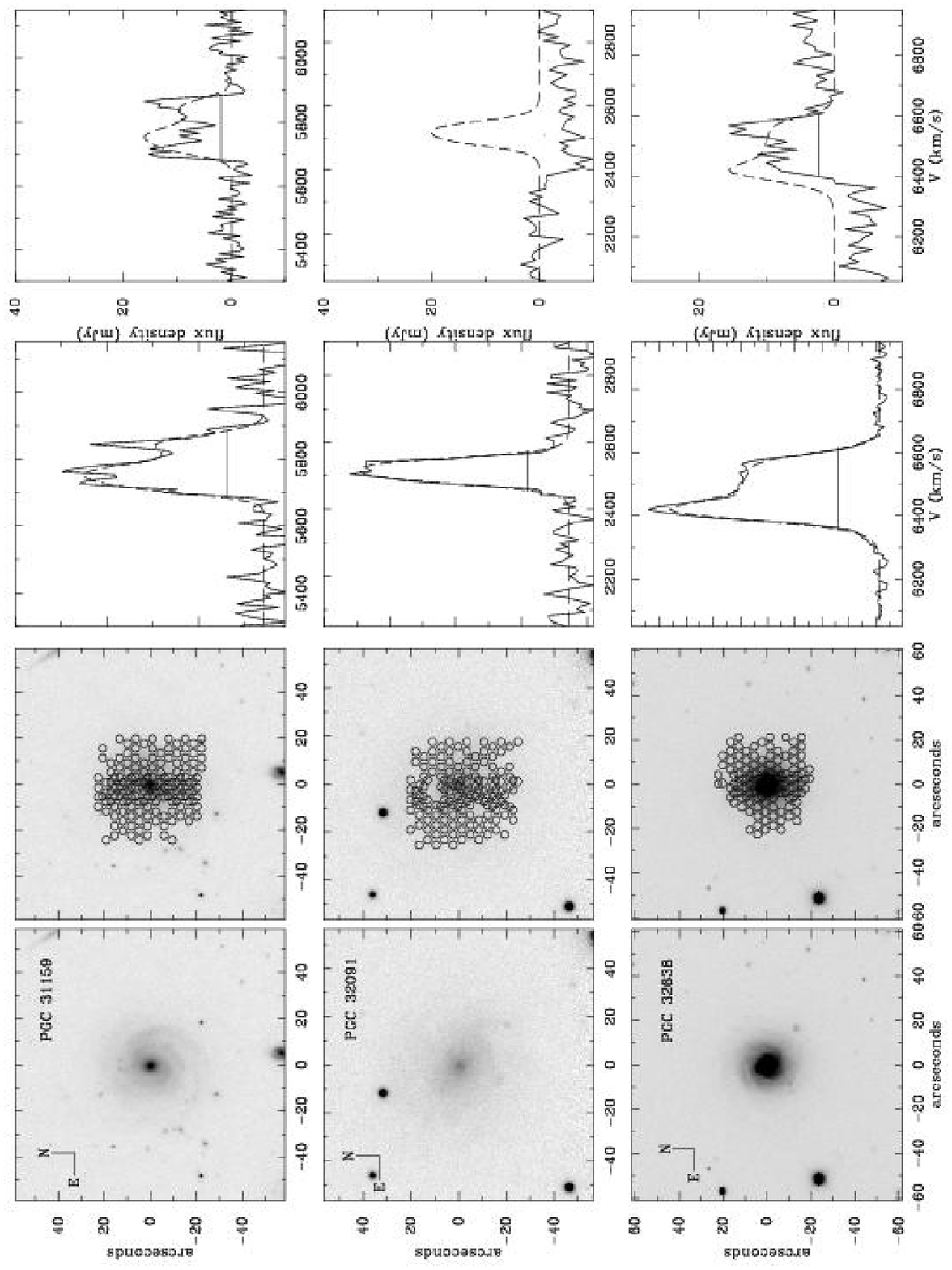}
\end{figure*}

\clearpage

\begin{figure*}
\vbox to 7.5in{\rule{0pt}{7.5in}}
\includegraphics{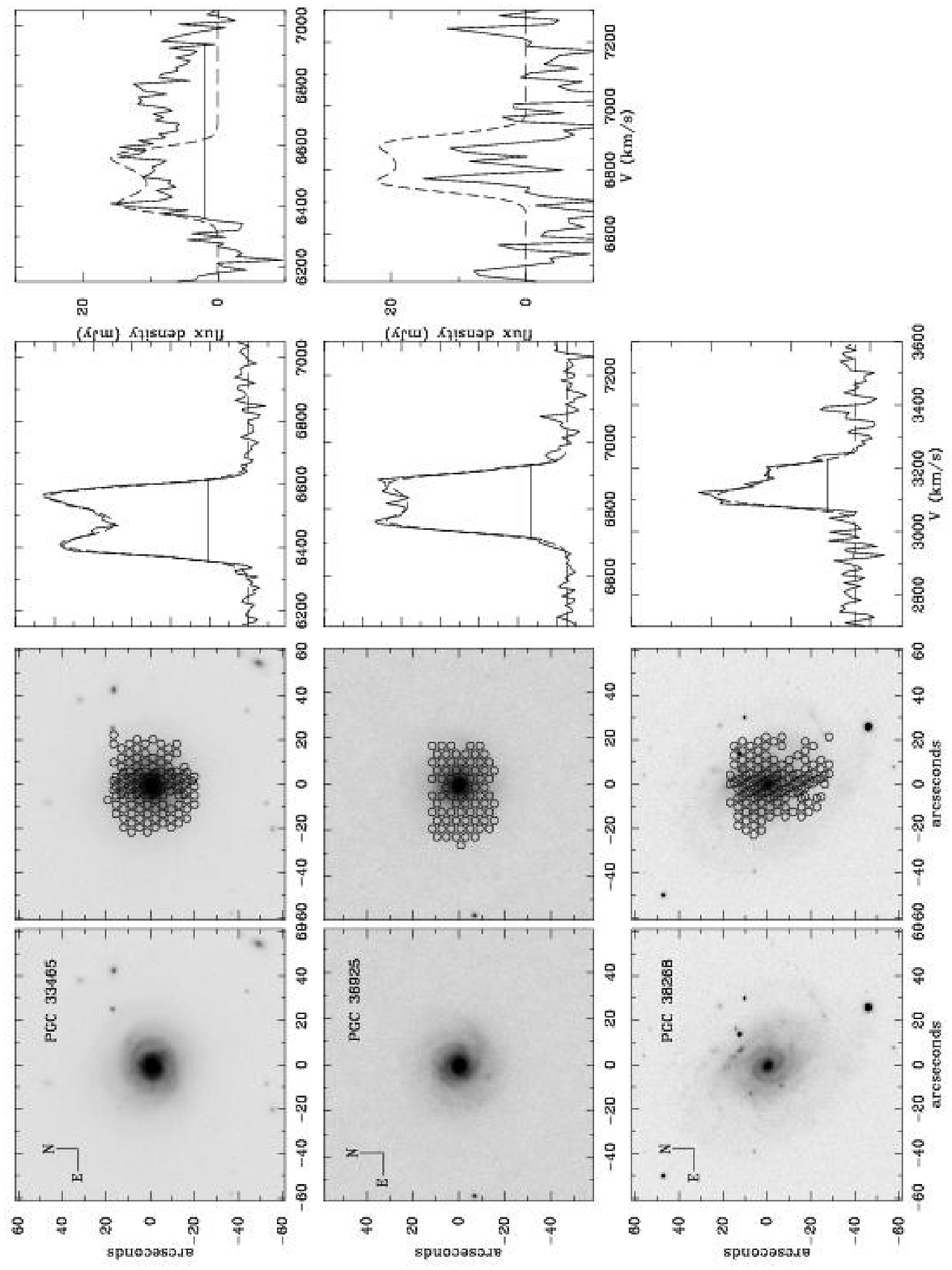}
\end{figure*}

\clearpage

\begin{figure*}
\vbox to 7.5in{\rule{0pt}{7.5in}}
\includegraphics{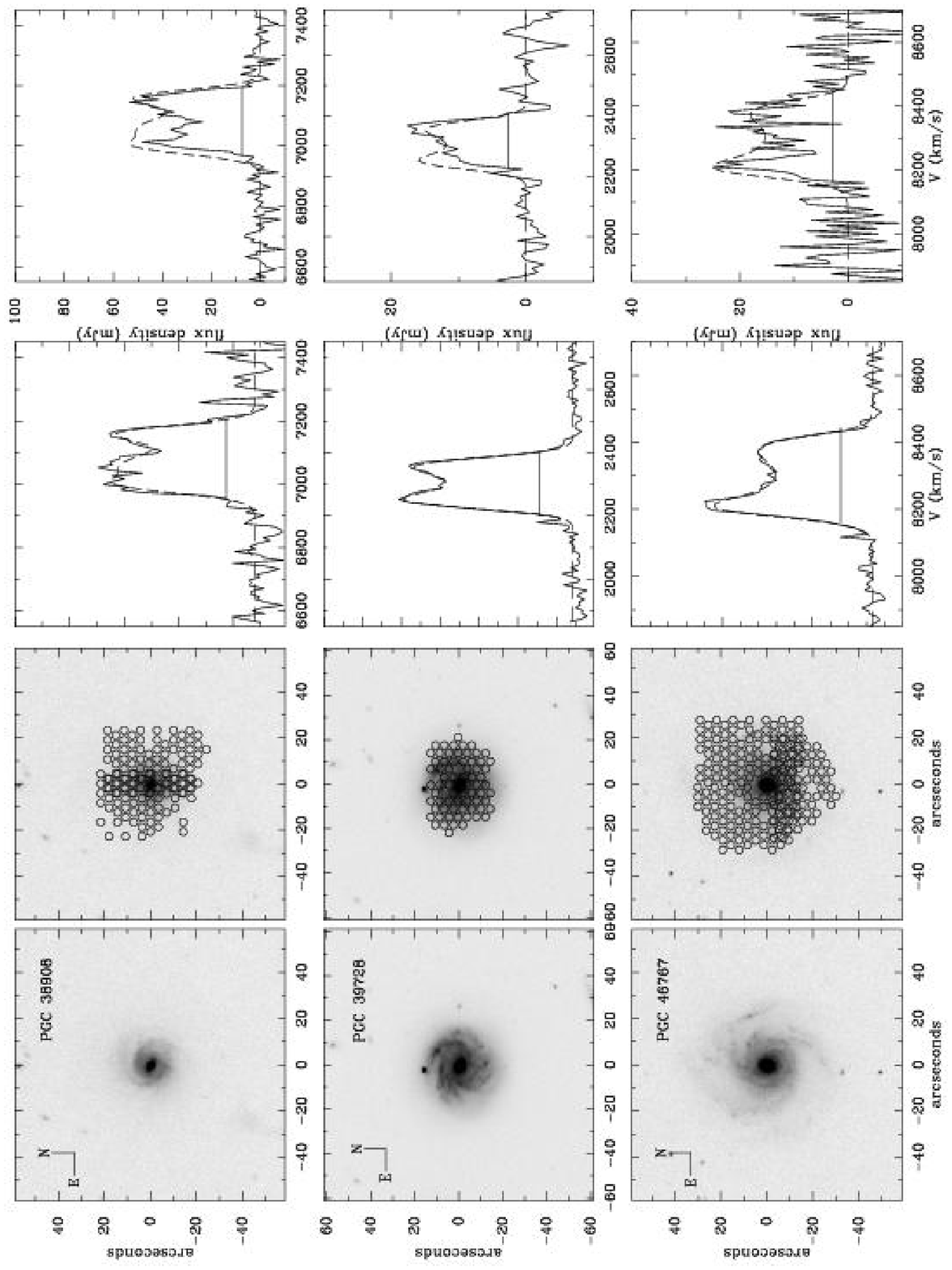}
\end{figure*}

\clearpage

\begin{figure*}
\vbox to 7.5in{\rule{0pt}{7.5in}}
\includegraphics{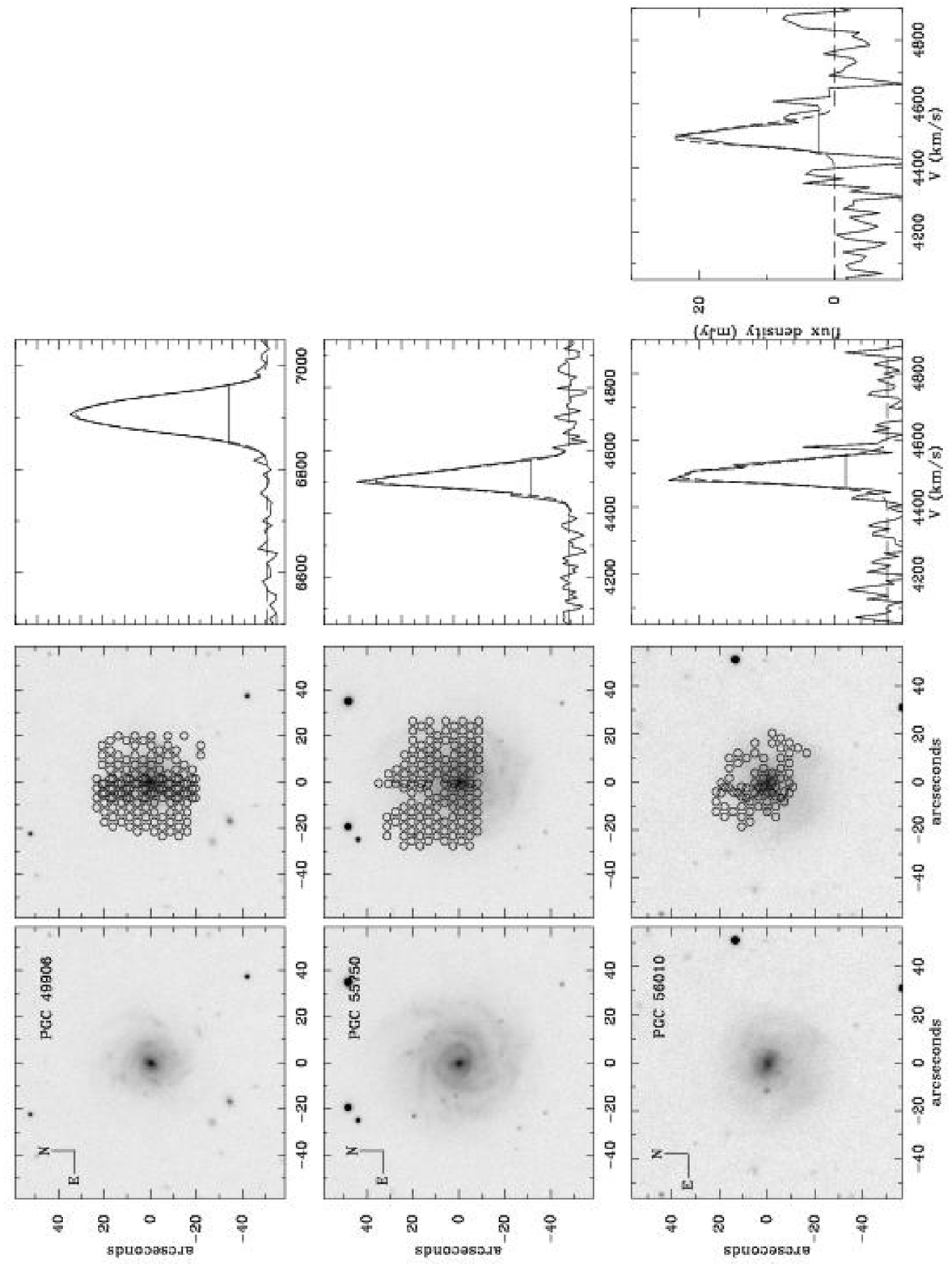}
\end{figure*}

\clearpage

\begin{figure*}
\vbox to 7.5in{\rule{0pt}{7.5in}}
\includegraphics{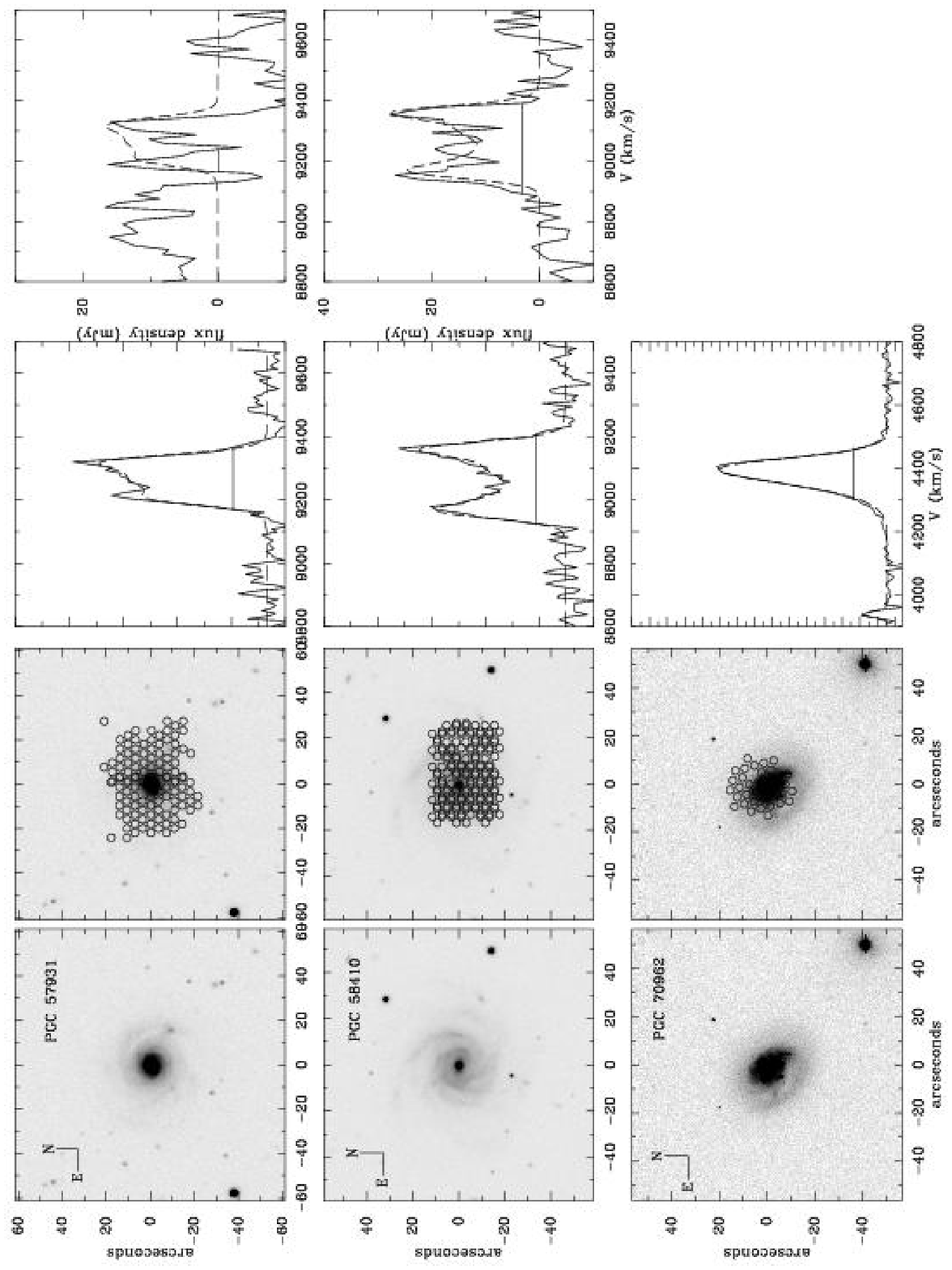}
\end{figure*}

\clearpage

\begin{figure*}
\vbox to 7.5in{\rule{0pt}{7.5in}}
\includegraphics{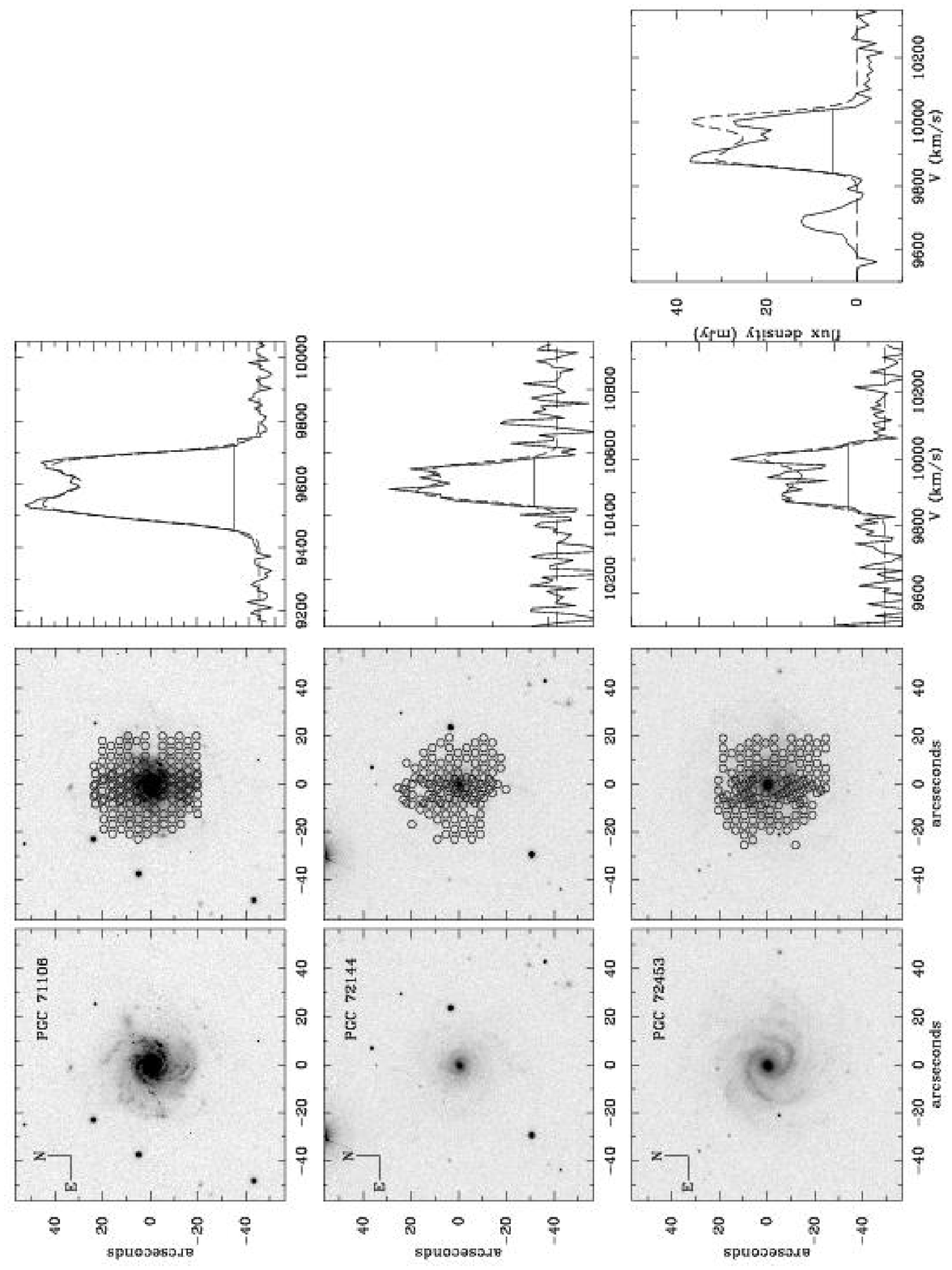}
\end{figure*}

\end{document}